\documentclass[fleqn,usenatbib]{mnras}

\usepackage{newtxtext,newtxmath}
\usepackage[T1]{fontenc}
\usepackage{ae,aecompl}
\usepackage{xtab,afterpage}
\usepackage{array,longtable}

\usepackage{graphicx}	% Including figure files
\usepackage{amsmath}	% Advanced maths commands
\usepackage{amssymb}	% Extra maths symbols
\title[Astrometric and photometric study of 7 Open Clusters]
{Astrometric and photometric study of Dias 4, Dias 6 and other five open clusters using ground based and Gaia DR2 data}

\author[Dias, W. S., et. al.]{
Dias, W., S.,$^{1}$\thanks{E-mail: wiltonsdias@yahoo.com.br}
Monteiro, H.,$^{1}$
L\'epine,J. R. D. ,$^{2}$
Prates, R.,$^{3}$
Gneiding, C., D.,$^{3}$
\newauthor Sacchi, M., $^{4}$
\\
$^{1}${Universidade Federal de Itajub\'a, Instituto de F\'isica e Qu\'imica, Itajub\'a-MG, Brazil}\\
$^{2}${Universidade de S\~ao Paulo, Instituto de Astronomia, Geof\'isica e Ci\^encias Atmosf\'ericas, S\~ao Paulo - SP, Brazil}\\
$^{3}${Laborat\'orio Nacional de Astrof\'isica, Rua Estados Unidos, 154, CEP 37504-364, Itajub\'a-MG - MG, Brazil}\\
$^{4}${Charles Sturt University, The Grange Chancellery, Panorama Avenue, Bathurst NSW Australia 2795}
}

\date{Accepted XXX. Received YYY; in original form ZZZ}

\pubyear{2017}

\begin{document}
\label{firstpage}
\pagerange{\pageref{firstpage}--\pageref{lastpage}}
\maketitle

\begin{abstract}
We present a study of 7 southern open clusters based on UBVRI CCD photomety (Johnsons-Cousins system) and Gaia DR2 data. Dias 4, Dias 6 and four other clusters had UBVRI photometric observations determined for the first time. From the observational UBVRI data we obtained photometric membership probability estimates and, using the proper motions from the UCAC5 catalog, we also determined the kinematic membership. From Gaia DR2 astrometric data we determine the stellar membership using proper motions and parallaxes, taking into account the full covariance matrix. For both independent sets of data and membership we apply our non subjective multidimensional global optimization tool to fit theoretical isochrones to determine the distance, age, reddening, metallicity and binary fraction of the clusters.  
The results of the mean proper motions, distances and ages are in agreement, but the ones obtained from Gaia DR2 data are more precise in both membership selection and estimated parameters.
In the case of NGC 6087, the Cepheid S Nor, member of the open cluster, was used to obtain an independent distance estimate, confirming the one determined by our fitting method. We also report a serendipitous discovery of two new clusters in the extended field near what was originally Dias 4.

\end{abstract}


%%%%%%%%%%%%%%%%%%%%%%%%%%%%%%%%%%%%%%%%%%%%%%%%%%

%%%%%%%%%%%%%%%%% BODY OF PAPER %%%%%%%%%%%%%%%%%%

\section{Introduction}

This paper is part of a series motivated by the need for a homogeneous set of open cluster parameters such as distances and ages, to improve the New catalog of Optically Visible Open Clusters and Candidates published by \cite{Dias2002} (hereafter DAML02\footnote{The latest version (3.5) can be accessed on line at \url{https://wilton.unifei.edu.br/ocdb/}.}) which is widely used in different astrophysical researches and specially by our group, to investigate the spiral arms structure \citep{Dias2005} and evolution of the galactic disk \citep{Lepine2011}.

A major problem in studies of open clusters is the determination of the membership probabilities of stars. Clearly with more accurate determination of cluster membership it is possible to better constrain estimates of distances, ages, and velocities (proper motion and radial velocity). Our team has contributed in this topic through membership probability estimates from both photometric \citep{Monteiro2017} and proper motion data \citep{Dias2014,Dias2018}. In this study we show how the solution of this question evolved in the era of the Gaia mission \citep{GAIA-DR22018}.

The open clusters investigated in the present study were selected from DAML02 with the aim to provide a set of homogeneous UBVRI data and parameters determined from them in a non subjective isochrone-fitting approach. For all the clusters studied here, except for NGC 6087, these are the first results based on UBVRI data. The cluster NGC 6087 was observed in this run because the Cepheid S Nor, member of the cluster, allows us to obtain an independent check of the distance estimate by our isochrone fit method.

Despite the fact that several of the clusters investigated in the present work had previous determinations of distances and ages, important corrections were needed given the discrepancies between results obtained by different works using different methods, as discussed by \citet{Netopil2015}. There are several reasons that may explain the differences in the parameters of the clusters published in the literature. However, the most important reasons are: 1) characteristics and quality of the data, 2) determination of the membership probabilities of the stars and 3) the isochrone-fitting method used.
In our previous works we emphasized some important points in determining accurately the parameters of open clusters. Two of the most important ones are the relevance of the U filter to correctly estimate the extinction (which is limited by the spectral-type/reddening degeneracy of the color-color diagram) and the subjectivity in the visual isochrone fits, which are still used in many works.

With the publication of the Gaia DR2 catalog, with its unprecedented astrometric precision, we can give answers to relevant questions in the study of open clusters, such as: how does data quality improve membership determination? What is the confidence and precision of the cluster parameters determined by the isochrone fits from Gaia DR2 photometric data alone?

The objects studied in this work, which were observed in our photometric Survey of Southern Open Clusters, are briefly presented in the next Section. In general, since the survey is a long period project, the clusters observed are selected based mainly on their visibility (given the allocated telescope time) as well as the most up to date information on the availability of good quality CCD UBVRI photometry for each object. This requirement is justified given our goal to work towards an homogeneous sample of objects. Some objects are re-observed if additional information can be gained, as in the case of the cluster NGC 6087 described below.

The available data is described in two main Sections, 3 and 4. 
Section 3 refers to UBVRI data, providing details of the observation and membership determination, and in Section 4 we present the Gaia DR2 data and astrometric membership. 
In Sec. 5 we describe the isochrone fitting method used. In Sec. 6 we present a discussion of the results obtained for each cluster. 
In Sec. 7 we compare the results with those published by different studies and discuss results obtained from UBVRI and Gaia DR2 data.
In Sec. 8 we give our final conclusions.

\section{Overview of the investigated open clusters}

\subsection{Collinder 307}

\citet{Moffat1975} concluded there was no evidence of the existence of a real open cluster based on UBV photoelectric photometry of 12 stars with $V \leq 12.4$. 

However, \citet{Giorgi2004} studied the cluster performing UBV CCD photometry and spectroscopy of the brighter stars in the field. Based on the results from the stars with $ V \leq 12 $ 
the authors estimated distance of 1740 pc and age between 250 and 400 Myr (logt between 8.40 and 8.60) for $E(B-V)=0.8$.

\citet{Carraro2006} performed CCD observations with V and I filters of a field of 13.5 arcmin centered on Collinder 307. The color-magnitude diagram (CMD), as also presented by \citet{Giorgi2004}, shows a clear feature of a real open cluster.
The parameters determined by the authors indicate an intermediate age open cluster (logt = 8.60 yr) at 1300 pc.

\citet{Kharchenko2013} (hereafter K13) and \citet{Bukowiecki2011} (hereafter B11) studied Collinder 307 based on infrared JH$K_{s}$ data from 2MASS catalog \citep{Skrutskie2006} but obtained discordant parameters. While K13 estimated distance of 2065 pc and a relatively younger age (logt = 7.25 yr), B11 found a distance of 1550 pc and logt = 8.4 yr.

\subsection{Dias 4}

Dias 4 was discovered in 2002 \citep{Dias2002} by visual inspection of the cluster field when constructing the DAML02 catalog.

It was studied by \citet{Tadross2009} and K13 using JH$K_{s}$ data from the 2MASS catalog. Both authors estimated distance of about 2100 pc and age of logt $\approx 9.2 yr$, for $E(B-V) \approx 0.5$ using different approaches for isochrone fit procedure assuming solar metal content.

\subsection{Dias 6}

The open cluster Dias 6 was also discovered in 2002 \citep{Dias2002} by visual inspection of the cluster field when constructing the DAML02 catalog. Although the object has been studied by two authors, the full UBVRI data was up to now unavailable.

Fundamental parameters of Dias 6 were published for the first time by \citet{Bica2004}. In that paper the object was assigned as Cluster 2. 
The authors used JH$K_{s}$ data from the 2MASS catalog to derive the fundamental parameters: $d = 2190$ pc, $E(B-V)=0.91$ mag and age log t$\approx$ 8.70 yr. 
The cluster was also studied by \citet{Tadross2008} who
also used data from 2MASS and obtained $E(B-V)=0.91$ mag, $\log{t}= 8.78$ yrs and a closer distance of about 1580 pc.

\subsection{IC 4651}

IC 4651 is a well-studied open cluster for which data in different photometric systems as well as spectroscopy is available. However, no UBVRI CCD data has been found in the literature.

The cluster's stars are well separated from the background stars and therefore presents the CMD with well-defined main sequence (MS), turn-off (TO) and red giant clump.
In the literature one can find several studies of the cluster and their stellar population (see \citet{Netopil2015}
and \citet{Mermilliod1995}). 
% In table \ref{tab:results} we show a compilation of the published parameters for this object obtained using different data sets and techniques.

\subsection{NGC 5138}

NGC 5138 was studied by UBV photoelectric data of some tens of stars by \citet{Lindoff1972}  and \citet{Claria1980}. Both authors have found $ E(B-V) \approx 0.25 $ but do not agree very well on distance and age values.
Using 2MASS data K13 estimate $E(B-V) = 0.312$ and a distance of 1819 pc in agreement with \citet{Claria1980}, but considering a slightly younger age.

To the best of our knowledge, there has been no UBVRI data for the stars of the NGC 5138 field. 

\subsection{NGC 6087}

This well-studied cluster was included in the observations because it contains a Cepheid (S Nor) which allowed us to verify independently that our method is producing reliable results. Well studied objects like these can also be used as important benchmarks when comparing different analysis techniques.

The cluster has been studied since the 1960s through optical data and more recently with near infrared 2MASS data. There is only one study based on CCD photometry with UBVRI data performed by \citet{Sagar1997}.
We refer the reader to the very recent study of \citet{Aidelman2018} for a complete compilation of the parameters of the cluster determined by different authors, using different data and techniques. 
% In order to facilitate it is reproduced in Table \ref{tab:results}.

\subsection{NGC 6178}

Despite being easily seen in the DSS image, NGC 6178 is a poorly studied open cluster as shown by the number of the published results compiled in Table \ref{tab:results}. 

At optical wavelengths, the first study was carried out by \citet{Moffat1973} who estimated $E(B-V) = 0.24$ and a distance of 910 pc using UBV photoelectric data of 19 stars. \citet{Piatti2000} using BVI CCD data estimated a similar $E(B-V)$ but a distance of 1300 pc with age $logt = 7.60 yr$. 
K13 using 2MASS data obtained $E(B-V) \approx 0.24$ and distance 1100 pc and logt = 7.510 yr. It is interesting to highlight the study of \citet{Rastorguev1999} which estimated a closer distance of 880 pc by statistical parallax.

In this work we present for the first time UBVRI CCD data of the stars in the field of NGC 6178.

The fundamental parameters published in the literature for the clusters discussed above are summarized in Table \ref{tab:results}.

\begin{table*}
  \caption{Parameters obtained for the investigated  clusters. In the columns two to six are given the parameters estimated in this work.
In columns seven, eight and nine previously published values are shown. In the first line of each cluster are given the parameters estimated from observed UBVRI data and in the last line the parameters estimated from Gaia data. The reference codes are detailed in the end of table together with the information of the filters used in the determined parameters.}
\label{tab:results} 
  \begin{tabular}{lccccccccccc}
    \hline\hline
              & \multicolumn{4}{c}{This work}                      & &\multicolumn{3}{c}{Literature} \\
    {Cluster}               &   {$E(B-V)$}     & {Distance}     & {Log(Age)}     & {Z}                 &  {Rv}               & {$E(B-V)$}         &  {Distance}        & {Log(Age)}      &  {Ref.} \\
    \hline      
Collinder~307               & $0.65\pm0.05$    & $1365\pm148 $  & $8.65\pm0.12$  & $0.008\pm0.007$     &   $3.04\pm0.08$     &   $0.67\pm0.04$    &  $1341\pm155   $   &    $8.50\pm0.17$&   II    \\ %erro = 50Myr
                            &                  &                &                &                     &                     &   $1.17\pm0.08$    &  $2065\pm227   $   &    $7.25\pm0.11$&   K13   \\ 
                            &                  &                &                &                     &                     &   $0.83\pm0.03$    &  $1551\pm107   $   &    $8.40\pm0.05$&   B11   \\ 
                            &                  &                &                &                     &                     &   $0.24\pm0.10$    &  $1300\pm227   $   &    $8.40\pm0.09$&   N     \\ 
Collinder~307               & $0.84\pm0.24$    & $1819\pm497 $  & $8.23\pm0.61$  & $0.021\pm0.006$     &   $3.24\pm0.41$     &                    &                    &                 &         \\              
                            
\hline
Dias~4                      & $0.36\pm0.04$    & $1342\pm 63 $  & $9.00\pm0.11$  & $0.018\pm0.005$     &   $3.09\pm0.06$     &   $0.60       $    &  $2150\pm100   $   &    $9.10       $&   T09   \\ 
                            &                  &                &                &                     &                     &   $0.58\pm0.04$    &  $2155\pm237   $   &    $9.26\pm0.04$&   K13   \\
                            &                  &                &                &                     &                     &                    &                    &                 &         \\
Dias~4a                     & $0.35\pm0.01$    & $1807\pm118 $  & $8.72\pm0.05$  & $0.021\pm0.006$     &   $3.51\pm0.30$     &                    &                    &                 &         \\ 
Dias~4b                     & $0.29\pm0.01$    & $1787\pm 46 $  & $7.53\pm0.08$  & $0.023\pm0.003$     &   $4.00\pm0.05$     &                    &                    &                 &         \\ 
\hline
Dias~6                      & $0.75\pm0.04$    & $2698\pm220 $  & $8.90\pm0.08$  & $0.018\pm0.008$     &   $2.82\pm0.02$     &   $0.91       $    &  $2190\pm      $   &    $8.70       $&   BB    \\
                            &                  &                &                &                     &                     &   $0.91       $    &  $1580\pm70    $   &    $8.78       $&   T08   \\
                            &                  &                &                &                     &                     &   $0.90\pm0.06$    &  $1916\pm211   $   &    $8.60\pm0.04$&   K13   \\
Dias~6                      & $0.88\pm0.03$    & $2627\pm182 $  & $9.02\pm0.04$  & $0.024\pm0.006$     &   $2.05\pm0.09$     &                    &                    &                 &         \\
\hline
IC~4651                     & $0.05\pm0.04$    & $ 965\pm 45 $  & $9.25\pm0.05$  & $0.025\pm0.006$     &   $2.85\pm0.06$     &   $0.15       $    &  $ 794         $   &    $     $      &   C     \\ 
                            &                  &                &                &                     &                     &   $0.12       $    &  $1150         $   &    $8.78 $      &   D     \\
                            &                  &                &                &                     &                     &   $0.086      $    &  $ 926         $   &    $9.38 $      &   A     \\
                            &                  &                &                &                     &                     &   $0.13\pm0.02$    &  $ 790\pm 60   $   &    $9.36   $     &   B     \\ 
                            &                  &                &                &                     &                     &   $0.10       $    &  $1010\pm 50   $   &    $9.23\pm0.04$   &   E     \\                              
                            &                  &                &                &                     &                     &   $0.09       $    &  $1072         $   &    $9.23\pm0.03$   &   TW    \\                              
                            &                  &                &                &                     &                     &   $0.12\pm0.01$    &  $ 888\pm 98   $   &    $9.25\pm0.11$&   K13   \\ 
                            &                  &                &                &                     &                     &   $0.15\pm0.03$    &  $1004\pm 59   $   &    $9.15\pm0.05$&   B11   \\ 
IC~4651                     & $0.06\pm0.02$    & $ 920\pm 29 $  & $9.31\pm0.06$  & $0.025\pm0.007$     &   $2.90\pm0.42$     &                    &                    &                 &         \\ 
\hline
NGC~5138                    & $0.18\pm0.03$    & $1422\pm156 $  & $8.10\pm0.16$  & $0.018\pm0.004$     &   $2.84\pm0.06$     &   $0.27       $    &  $1450         $   &    $8.70  $     &   D     \\
                            &                  &                &                &                     &                     &   $0.25\pm0.02$    &  $1800\pm180   $   &    $7.72  $     &   F     \\ 
                            &                  &                &                &                     &                     &   $0.31\pm0.02$    &  $1819\pm200   $   &    $7.55\pm0.11$&   K13  \\ 
NGC~5138                    & $0.20\pm0.02$    & $1857\pm 95 $  & $8.18\pm0.17$  & $0.027\pm0.004$     &   $3.97\pm0.04$     &                    &                    &                 &         \\
\hline
NGC~6087                    & $0.21\pm0.03$    & $1009\pm97 $   & $7.75\pm0.25$  & $0.014\pm0.004$     &   $2.95\pm0.10$     &   $0.22       $    &  $ 759\pm 10   $   &    $7.30       $&   I     \\ 
                            &                  &                &                &                     &                     &   $0.20       $    &  $ 912\pm 10   $   &    $7.30       $&   J     \\ 
                            &                  &                &                &                     &                     &   $0.20\pm0.01$    &  $ 871         $   &    $           $&   L     \\ 
                            &                  &                &                &                     &                     &   $0.18\pm0.01$    &  $ 832\pm 10   $   &    $           $&   M     \\ 
                            &                  &                &                &                     &                     &   $0.19\pm0.01$    &  $ 903\pm 11   $   &    $           $&   K     \\ 
                            &                  &                &                &                     &                     &   $0.22\pm0.03$    &  $ 960\pm130   $   &    $7.81  $     &   G     \\
                            &                  &                &                &                     &                     &   $0.13\pm0.03$    &  $ 851\pm 10   $   &                 &   H     \\ 
                            &                  &                &                &                     &                     &   $0.27\pm0.02$    &  $ 889\pm 98   $   &    $7.95\pm0.11$&   K13   \\ 
                            &                  &                &                &                     &                     &   $0.35\pm0.03$    &  $ 629\pm 54   $   &    $7.74       $&   Q     \\ 
NGC~6087                    & $0.20\pm0.02$    & $ 913\pm64 $   & $7.84\pm0.13$  & $0.017\pm0.006$     &   $3.87\pm0.13$     &                    &                    &                 &         \\ 
\hline
NGC~6178                    & $0.20\pm0.01$    & $893\pm105 $  & $7.85\pm0.44$  & $0.018\pm0.007$     &   $3.19\pm0.02$      &   $0.24       $    &  $ 910         $   &                 &   O     \\
                            &                  &                &                &                     &                     &   $0.20\pm0.05$    &  $1259\pm 11   $   &    $7.67       $&   P     \\ 
                            &                  &                &                &                     &                     &   $0.24\pm0.02$    &  $1011\pm111   $   &    $7.51\pm0.11$&   K13   \\ 
NGC~6178                    & $0.24\pm0.02$    & $817\pm108 $  & $7.30\pm0.16$  & $0.018\pm0.007$     &   $3.84\pm0.20$      &                    &                    &                 &         \\
\hline      
  \end{tabular} 
  \begin{flushleft}
  \tiny
A = \citet{Twarog1988} The E(B-V) and distance were based on uvby photometry. The age was estimated using CCD and photographic B and V filters.  \\
B = \citet{Kjeldsen1991}. Based on UBV CCD photometry.\\
C = \citet{Eggen1971}. Based on UBV photoeletric photometry.\\
D = \citet{Lindoff1972}. Based on UBV photoeletric photometry.\\
E = \citet{Meibom2002}. Based on CCD Stromgren photometry.\\
F = \citet{Claria1980}. Based on UBV photoeletric photometry and DDO photometry.\\
G = \citet{Sagar1997}. Based on UBVRI CCD photometry. \\
H = \citet{An2007}. Based on BVI CCD photometry and JH$K_{s}$ data from 2MASS catalog.\\
I = \citet{Fernie1961}. Based on UBV photoeletric photometry.  \\
J = \citet{Landolt1964}. Based on BV photoeletric and photographic photometry.\\
K = \citet{Turner1986}. Based on UBV photoeletric photometry.  \\
L = \citet{Breger1966}. Based on UBV Photoelectric photometry.\\
M = \citet{Schmidt1980}. Used V photoeletric, Stromgren b and y photometry, Washington m and c1 and $H\beta$. \\
N = \citet{Carraro2006}. Based on VI CCD photometry.\\
O = \citet{Moffat1973}. Based on UBV photoeletric photometry.\\
P = \citet{Piatti2000}. Based on BVI CCD photometry.\\
K13 = \citet{Kharchenko2013}. Based on data from 2MASS catalog.\\
B11 = \citet{Bukowiecki2011}. Based on data from 2MASS catalog.\\
T08 = \citet{Tadross2008}. Based on data from 2MASS catalog. \\
T09 = \citet{Tadross2009}. Based on data from 2MASS catalog. \\
BB = \citet{Bica2004}. Based on data from 2MASS catalog.\\
TW = \citet{Twarog2000}. Based on $uvby$H$\beta$ CCD photometry\\
II = \citet{Monteiro2017}. Based on UBVRI CCD photometry.\\
Q =  \citet{Aidelman2018}. Based on low-resolution spectra of B-type stars.\\
\end{flushleft}
\end{table*}

\section{UBVRI data}

\subsection{Observations and data reduction}

The observations were carried out using the B\&C 0.6m telescope at Pico dos Dias Observatory (LNA-Brazil) on the nights of July 05, 06 and 07 2013 under photometric conditions. 
The main detector of the survey was a SITe 1024$\times$1024 back-illuminated CCD with the plate scale of 0.6 arcsec pixel $^{-1}$, and the square field of ten arcminutes approximately and a Johnson-Cousins UBVRI filter set was employed.
The setup, configurations and observational strategy follow the procedures described in \citep{Caetano2015} (Paper I) and \citep{Monteiro2017} (Paper II). 
All images were collected in photometric stable conditions with seeing stable at 1.5 arcsec. 

We used the pipeline developed by our group to perform the standard data reduction process.
% Briefly, the images were bias and flat-field corrected and a mask was used to correct both flat-field and object frames 
% from shutter timing effects, prior to flat-field correction. 
In this run sky flats were also obtained.
We refer the reader to the Paper I for a complete description since the same procedure was applied in this work.

The instrumental magnitudes were determined by the point spread function (PSF) method
using the software STARFINDER \citep{Diolaiti2000}\footnote{\url{http://www.bo.astro.it/\~{}giangi/StarFinder/index.htm}}, executed automatically in the pipeline. 

At the end of the process, the equatorial coordinates obtained for each detected object allow us to cross-match the stars in each UBVRI filter.

\subsection{Transformation to the standard system}

Relations between the instrumental magnitudes and the magnitudes in the standard system were taken to be of the form of Eqs. \ref{eq:transformation}.\par

\begin{eqnarray}
\label{eq:transformation}
u = U + u1 + u2  X + u3 (U-B)\\
b = B + b1 + b2  X + b3 (B-V)\\
v = V + v1 + v2  X + v3 (B-V)\\
r = R + r1 + r2  X + r3 (V-R)\\ 
i = I + i1 + i2  X + i3 (V-I)
\end{eqnarray}
where upper case letters represent the magnitudes and colors in the standard system and lower case letters were adopted for the instrumental ones and X is the airmass. The coefficient values are reported in an online Table available at the web-site of the
project\footnote{\url{https://wilton.unifei.edu.br/OPDSurvey.html}}.

The best fit was obtained by a global optimization procedure minimizing the differences between the magnitudes of the observed standard stars and the cataloged values from \citet{Clem2013}. 
The final values of the coefficients are given by the mean of the results of one hundred runs using typically hundreds of stars. The standard deviation was used to represent the errors of the coefficients.  In this process the used parameter space limits were selected from the values of typical coefficients for the site of the OPD observatory assuming a range of $\pm 10\%$ to accommodate different kinds of variations and errors.  The \textit{rms} of the fits for the each night is typically lower than 0.03 in each filter.

The observed standard fields were TPheB, TPheG, SA110, WD1056-384, WD1153-484, LSE44, LSE259, MCT2019-4339, JL82, JL163, which amounts to hundreds of standard stars observed at different airmasses throughout each night.

For each studied cluster we give the catalog of calibrated data which includes our UBVRI photometry and error estimates, equatorial coordinates, proper motion and JH$K_{s}$ magnitudes, where the last two data were obtained from UCAC5 \citep{Zacharias2017} and 2MASS catalog, respectively.
The data provided in electronic tables and interesting plots as the residual of the fit to the standard stars and the errors as a function of the magnitude of all observed stars are given in the web-site of the project and vizier.

\subsection{Membership determination from UBVRI data and UCAC5 proper motions}

The fundamental assumption made in the membership determinations is that the members of a cluster are at the same distance, have the same spatial velocity, age and chemical composition due to being formed at the same time in the same environment. The assumption implies common proper motions, radial velocity, and apparent cluster sequence in the color-magnitude diagram. Given that, the astrometric data (parallaxes and proper motions) and radial velocity are important constraints that help to separate the population of the open cluster and field stars which are always present. 

The membership determination is intended to minimize the subjectivity in the selection of stars and to improve the isochrone fitting procedure since it can maximize the contrast of cluster features in relation to the field stars in a given CMD.

In \cite{Dias2012} our group presented a non parametric technique to estimate the membership of stars in the field of the open clusters from photometric data. It was improved by \cite{Monteiro2017} with the update of the code for performing isochrone fits based on the cross-entropy global optimization algorithm \footnote{The source code is available at \url{https://github.com/hektor-monteiro/OC_CEfit/releases}}.

In this work we used the same procedures defined in Paper II to determine the membership of the stars in the field of each studied clusters.
The membership considers, for each star, the position in the field, the spatial star density in that position and the star density in the multidimensional magnitude space, taking into account the photometric errors in each filter. The procedure works essentially as a multivariate kernel density estimate and we define the value obtained in this way  for each star as its photometric membership $P_{ph}$.

We also determined the membership of the stars in the region of each cluster using proper motion data from UCAC5 catalog \citep{Zacharias2017}. The method described below is the same we used in \citet{Dias2014} (hereafter D14) and \citet{Dias2018}. 

The model considers that the distribution of proper motions of the stars in a cluster's region can be represented by two elliptical bivariate Gaussians following \citet{Zhao1990}, including the proper motion's errors in the frequency function. The expressions used are presented in Eq.\ref{eq:cluster} and Eq.\ref{eq:field}.
The notation is \textit{c} and \textit{f} subscripts for cluster and field parameters, respectively, x for the coordinate
$\mu_{\alpha} \cos \delta$, and y for the coordinate
$\mu_{\delta}$.
$\Phi = \Phi_c + \Phi_f$ is the total probability distribution,
$(\mu_{x,c},\mu_{y,c})$ are the averages of the cluster distribution with
standard deviations $\sigma_{x,c}$ and $\sigma_{y,c}$,
$(\mu_{x,f},\mu_{y,f})$ are the averages of the field distribution with
standard deviations $\sigma_{x,f}$ and $\sigma_{y,f}$, and $\rho_{c}$
and $\rho_{f}$ are the correlation coefficients of cluster and field
stars. The values $(\mu_{x},\mu_{y})$ are the component of the proper motion of each star, and
$\epsilon_{i}$ is the formal error in proper motion given by the catalog.

\begin{eqnarray}
\label{eq:cluster}
\Phi_c(\mu_{x},\mu_{y}) =
  \frac{n_{c}}{2\pi\sqrt{\sigma_{x,c}^2+\epsilon_{i}^2}
    \sqrt{\sigma_{y,c}^2+\epsilon_{i}^2}
    \sqrt{1-\rho_{c}^2}} . . \\
. . X \exp\{ -\frac{1}{2(1-\rho_{c}^2)} [ 
\frac{(\mu_x - \mu_{x,c})^2}{\sigma_{x,c}^2+\epsilon_{i}^2}  . . \nonumber
\\ \  + . .
\frac{(\mu_y - \mu_{y,c})^2}{\sigma_{y,c}^2+\epsilon_{i}^2}
-2\rho_{c}(\frac{\mu_x-\mu_{x,c}}{\sqrt{\sigma_{x,c}^2+\epsilon_{i}^2}})
(\frac{\mu_y-\mu_{y,c}}{\sqrt{\sigma_{y,c}^2+\epsilon_{i}^2}})
] \}\ \ ,
\nonumber
%\label{eq:cluster}
\end{eqnarray}

\begin{eqnarray}
\label{eq:field}
\Phi_f(\mu_x,\mu_y) =
  \frac{1-n_{c}}{2\pi\sqrt{\sigma_{x,f}^2+\epsilon_{i}^2}
    \sqrt{\sigma_{y,f}^2+\epsilon_{i}^2}
    \sqrt{1-\rho_{f}^2}} . . \\
. . X \exp\{ -\frac{1}{2(1-\rho_{f}^2)} [ 
\frac{(\mu_x - \mu_{x,f})^2}{\sigma_{x,f}^2+\epsilon_{i}^2}  . . \nonumber
\\ \  + . .
\frac{(\mu_y - \mu_{y,f})^2}{\sigma_{y,f}^2+\epsilon_{i}^2}
-2\rho_{f}(\frac{\mu_x-\mu_{x,f}}{\sqrt{\sigma_{x,f}^2+\epsilon_{i}^2}})
(\frac{\mu_y-\mu_{y,f}}{\sqrt{\sigma_{y,f}^2+\epsilon_{i}^2}})
] \}\ \ 
\nonumber
%\label{eq:field}
\end{eqnarray} 

  The probability density function for the whole sample is simply given by Eq.\ref{eq:probability}, where
{$n_c$ and $n_f$ are the number of cluster and field stars (non-members), respectively, normalized with respect to the total number of stars in the field.

\begin{eqnarray}
%\begin{equation}
\Phi(\mu_x,\mu_y) = n_c \Phi_c(\mu_x,\mu_y) +
n_f \Phi_f(\mu_x,\mu_y)\ \ 
\label{eq:probability}
%\nonumber
%\end{equation}
\end{eqnarray}

With the frequency function parameters we determined
the individual probability of the membership of each \textit{ith}-star in the cluster by $P_{i} = \Phi_{c_{i}}/\Phi_{i}$.
%In this work we used this procedure applied to the UCAC5 stellar proper motion \citep{Zacharias2017} to estimate the membership probability of each star in the observed cluster region. 
Hereafter we call this membership probability as $P_{\mu}$. 

In Table \ref{tab:kin} we present the parameters provided by the
method described below, with the mean proper motions of the
clusters ($\mu_{\alpha} \cos \delta$ = $\mu_{x,c}$ , $\mu_{\delta}$ = $\mu_{y,c}$ and standard deviations $\sigma$) and the number of cluster members estimated. The two-dimensional obtained Gaussian fit to the population of field stars and the number of field stars, the
orientation angle of the minor axis of elliptical field star proper motion distribution are also given. In the second column of Table \ref{tab:kin} the radius (centered on the coordinates
of the cluster obtained from DAML02) used to extract the stars from the UCAC5 catalog is given. 

As the photometric membership ($P_{ph}$) and membership from proper motions ($P_{\mu}$) are independent of each other we derived a combined probability P as the mean of $P_{ph}$ and $P_{\mu}$. 
In this work we consider the stars with $P \geq 51\%$ as members of the cluster.
We use P to weight the isochrone fit in our code. For each cluster, a table with the value of P is given in electronic format in the web-site of the project. 

\begin{table*}
\caption[]{Results of mean proper motion and dispersion obtained using the UCAC5 stellar proper motion.  
The meaning of the
symbols are as follows:
R is the radius (in arcmin) used for each cluster to extract the UCAC5 data, centered on the coordinates of the cluster obtained from DAML02. 
$N_{c}$ is the number of cluster stars;
$N_{f}$ is the number of field stars;
$\mu_{\alpha}cos{\delta}$ and $\mu_{\delta}$ are the  proper motion components in mas yr$^{-1}$;
$\sigma$ is the dispersion of cluster stars' proper motions;
$\sigma\mu_{\alpha}cos{\delta}$ and $\sigma \mu_{\delta}$  
are the dispersions of the components of the field stars' proper motions;
$\rho$ is the orientation angle of the minor axis of the elliptic
proper motion distribution.}
\label{tab:kin}
\begin{center}
\begin{tabular}{lccccccc|ccccccc}
\hline \hline
&\multicolumn{7}{c|}{Cluster} & \multicolumn{4}{c}{Field}\\
cluster           &         R   &    $N_{c}$   & $\mu_{\alpha}cos{\delta}$&    $\sigma\mu_{\alpha}cos{\delta}$  &      $\mu_{\delta}$  &     $\sigma \mu_{\delta}$  &     $\rho_{c}$   &    $N_{f}$  &    $\mu_{\alpha}cos{\delta}$   &     $\sigma\mu_{\alpha}cos{\delta}$   &       $\mu_{\delta}$  &    $\sigma \mu_{\delta}$   &      $\rho_{f}$  \\
Collinder~307     &      4.00   &      196   &      -1.86   &     0.24  &     -3.77  &      0.14  &     0.15   &       82  &   -6.16   &      6.66   &      -8.25  &     6.41   &      0.31  \\
Dias~4            &      4.70   &      155   &      -5.70   &     1.09  &     -2.21  &      0.10  &     0.01   &       75  &   -9.36   &      7.95   &      -5.57  &     6.08   &      0.18  \\
Dias~6            &      4.50   &      175   &       0.80   &     0.66  &     -1.79  &      1.32  &     0.18   &       41  &   -1.19   &      4.55   &      -3.12  &     6.57   &     -0.23  \\
IC~4651           &      6.50   &      552   &      -2.49   &     0.76  &     -5.15  &      0.13  &     0.08   &      253  &   -1.83   &      5.68   &      -4.91  &     5.92   &      0.17  \\
NGC~5138          &      5.00   &      351   &      -5.92   &     2.73  &     -2.09  &      1.12  &    -0.02   &      124  &   -9.50   &     10.01   &      -3.60  &     8.67   &      0.11  \\
NGC~6087          &      8.50   &      972   &      -2.85   &     2.25  &     -4.38  &      2.15  &     0.27   &      241  &   -4.43   &      7.24   &      -6.83  &     8.92   &      0.00  \\
NGC~6178          &      4.00   &      287   &      -1.19   &     0.93  &     -3.03  &      1.50  &     0.00   &       94  &   -4.38   &      6.67   &      -5.87  &     9.35   &      0.04  \\
\hline
\end{tabular}
\end{center}
\end{table*}

\section{Gaia DR2 data}

In this work we also use Gaia DR2 catalog that presents all sky astrometric and photometric data with unprecedented precision for more than one billion objects with Gmag less than 21. All details of the catalog as well as instructions on its correct use are provided in \citet{GaiaDR2-Brow2018} and \citet{Gaia-DR2-Luri2018}.

We searched for Gaia DR2 stars in the sky area of the selected clusters, using the central coordinates and the apparent diameters taken from the DAML02 Catalog. To include virtually all possible members of the studied clusters, we opted to use radius 2 arcmin bigger than those provided by DAML02 catalog. In Table \ref{tab:meanparams} we give the radius used for each cluster to extract the Gaia DR2 data. 

A direct consequence of the quality of the Gaia DR2 astrometric data is the possibility of determining better membership of stars of the open clusters. 
In a vector proper motion diagram (VPD) of stars in a region of a open cluster, using Gaia data, it is usually possible to see clearly the expected clump which is due to the stars of the cluster presenting the same spatial velocity, that is, similar proper motion, parallax and radial velocity.
This precision of the astrometric data not only allows a better distinction of the cluster in the VPD, but also opens new windows in the analysis and interpretation of the stars in cluster fields, as presented for the case of the open cluster Dias 4 discussed in the Sec. 5.

\subsection{Membership  determination from Gaia DR2 astrometric data}

From the statistical point of view, the clustering of the stars in space comes together with clustering in the data space of proper motion plus parallax. Therefore, the closer a given star is to the density peak
in that space, the higher the probability of the star being a member. In the space formed by the proper motion and parallax ($\mu_{\alpha}cos{\delta}$, $\mu_{\delta}$, $\varpi$), the problem can be formulated as a maximum likelihood one, based on the assumption of normally distributed uncertainties in proper motion and parallax as given in the equation \ref{eq:multinormal} below.

\begin{equation} 
\label{eq:multinormal}
f(\mathbf {X}) = {\frac {\exp \left(-{\frac {1}{2}}({\mathbf {X} }-{\boldsymbol {\mu }})^{\mathrm {T} }{\boldsymbol {\Sigma }}^{-1}({\mathbf {X} }-{\boldsymbol {\mu }})\right)}{\sqrt {(2\pi )^{k}|{\boldsymbol {\Sigma }}|}}}
\end{equation}

where $\mathbf {X}$ is the column vector ($\mu_{\alpha}cos{\delta}$, $\mu_{\delta}$, $\varpi$) composed of the proper motion components and the parallax, $\boldsymbol {\mu }$ the mean column vector and $|{\boldsymbol {\Sigma }}|$ is the full covariance matrix which incorporates all uncertainties ($\sigma$) and their correlations ($\rho$) given in the Gaia DR2 catalog as given in the equation \ref{eq:covariancematrix}.  

        \begin{equation} \label{eq:covariancematrix}
        \mathrm {\boldsymbol {\Sigma }} = 
                 \begin{bmatrix}
                        \sigma^{2}_{\mu_{\alpha}*}      & \sigma_{\mu_{\alpha}*} \sigma_{\mu_{\delta}} \rho_{\mu_{\alpha}*\mu_{\delta}}          & \sigma_{\mu_{\alpha}*} \sigma_{\varpi} \rho_{\mu_{\alpha}*\varpi}      \\
                        \sigma_{\mu_{\alpha}*} \sigma_{\mu_{\delta}} \rho_{\mu_{\alpha}*\mu_{\delta}}   &  \sigma^{2}_{\mu_{\delta}}      &  \sigma_{\mu_{\delta}} \sigma_{\varpi} \rho_{ \mu_{\delta} \varpi}     \\
                        \sigma_{\mu_{\alpha}*} \sigma_{\varpi} \rho_{\mu_{\alpha}*\varpi}       &  \sigma_{\mu_{\delta}} \sigma_{\varpi} \rho_{ \mu_{\delta} \varpi}      & \sigma^{2}_{\varpi}     \\
                 \end{bmatrix} ,
        \end{equation}

where $\mu_{\alpha}*$ = $\mu_{\alpha}cos{\delta}$.  
% \noindent where $\sigma_{\mu_{\alpha}*}$, $\sigma_{\mu_{\delta}}$, and $\sigma_{\varpi}$,
% and $\rho_{\mu_{\alpha}*\mu_{\delta}}$, $\rho_{\mu_{\alpha}*\varpi}$, and $\rho_{ \mu_{\delta} \varpi}$ (the correlation coefficients) given in the catalog. 

Since the proper motion and parallaxes were determined from a simultaneous five-parameter fit of an astrometric source model to the data, they present an uncertainty derived from the formal errors and correlation coefficients between the estimated parameters.   
So, as described in \citet{Gaia-DR2-Luri2018} the estimated model parameters have to take into account the uncertainty of the measurements as well as their correlations. We follow the recommendations given by \citet{Gaia-DR2-Luri2018} using the diagonal elements (the standard uncertainties) and the off-diagonal elements (the correlations between the uncertainties) in the model given by the Eq. \ref{eq:multinormal}.

In Table \ref{tab:meanparams} we present the mean astrometric parameters determined from the method described in this Section. A table available at the web-site of the project gives the value of P for each star in the field of the cluster.  

\begin{table*}
\caption{Mean astrometric parameters from Gaia DR2 data.
R is the radius used for each cluster to extract the Gaia DR2 data, 
centered on the coordinates RA and DEC.}
\label{tab:meanparams}
\begin{tabular}{lccccccc}
\hline
\noalign{\smallskip}
Cluster & RA  &   DEC  &  R  &$ \varpi $   &   $\mu_{\alpha}cos{\delta}$  &   $ \mu_{\delta} $   &  Nc\\ 
        & deg &   deg  &  arcmin  &mas          &   mas yr$^{-1}$      &   mas yr$^{-1}$      &  \\
\hline
\noalign{\smallskip}
Collinder~307  & 248.807175     &   -51.019264    &  4.5   & $0.505\pm0.033$     & $-0.778\pm0.040$   & $-3.319\pm0.030$     &  484  \\
Dias~4         & 205.887393     &   -63.040776    &  5.2   & $0.528\pm0.023$     & $-4.428\pm0.048$   & $-2.315\pm0.096$     &  200  \\
Dias~6         & 277.611861     &   -12.329109    &  5.0   & $0.321\pm0.050$     & $ 0.478\pm0.015$   & $-0.573\pm0.020$     &  953  \\
IC~4651        & 261.050536     &   -49.964520    &  7.0   & $1.062\pm0.047$     & $-2.406\pm0.021$   & $-5.031\pm0.029$     &  814  \\
NGC~5138       & 201.881011     &   -59.032486    &  5.5   & $0.518\pm0.049$     & $-3.550\pm0.015$   & $-1.402\pm0.005$     &  867  \\
NGC~6087       & 244.559239     &   -58.036557    &  9.0   & $1.018\pm0.048$     & $-1.535\pm0.054$   & $-2.387\pm0.024$     &  1180  \\
NGC~6178       & 248.874097     &   -45.593121    &  4.5   & $1.112\pm0.050$     & $ 0.536\pm1.147$   & $-3.272\pm0.437$     &  380  \\
\hline
\noalign{\smallskip}
\end{tabular}
\end{table*}

\section{Isochrone fit using the cross-entropy method}
After determining the membership probabilities from the astrometric data we applied the cross-entropy (CE) method to fit the theoretical isochrones to the multidimensional color-magnitude diagrams.

Basically, the CE method, which has been described in detail in our previous papers, involves an iterative statistical procedure where the following is done in each iteration:

\begin{itemize}
\item random generation of the initial sample of fit parameters,
  respecting predefined criteria;
\item selection of the best candidates based on calculated weighted
  likelihood values;
\item generation of random fit parameter sample derived from a
  new distribution based on the previous step;
\item repeat until convergence or stopping criteria reached.
\end{itemize}

The code uses the tabulated isochrones from \citet{Bressan2012} and search for the solutions in the parameter space defined as follows:
\begin{itemize}
\item age: from log(age) =6.60 to log(age) =10.15;
\item distance: from 1 to 10 000 parsecs;
\item $E(B-V)$: from 0.0 to 3.0; 
\item Metallicity (Z): from 0.001 to 0.30 dex with steps of $Z = 0.05$ dex.
\item $R_{v}$: from 2.8 to 3.2
\end{itemize}

In Table \ref{tab:ceparameters} we present the structural cluster parameters used in the isochrone fit applied to the observations data. The equatorial coordinates $\alpha_{c}$ and $\delta_{c}$ used as center coordinates of the cluster in the observed field were obtained as the point where the density
  distribution peaks. The characteristic cluster radius ($R_{c}$) used was calculated based on the the standard deviation of all the distances between stars in the sample and the center coordinate $\alpha_{c}$ and $\delta_{c}$. The density distribution of stars ($\rho$) is obtained from the $\alpha$ and $\delta$ coordinates using a Gaussian kernel. The kernel used has a standard deviation determined using the standard deviation of the measured distances between stars in the sample and applying the well known Silverman's rule, such that  ${SIG\_PIX}   \approx 1.06 \sigma N^{-1/5}$, where $\sigma$ is the standard
  deviation of distances and N is the number of stars in the sample. 
We use the characteristic cluster radius at 1$\sigma$. 
The other tunning parameters of the method are adopted as described in detail in \citet{Monteiro2017} and references therein. For the clusters Dias 6, NGC 6087 and NGC 6178 we opted to use the cluster region defined by an iso-density limit obtained from the density map. The density limit is defined as the density value 1$\sigma$ above the mean field density.
  
To determine the final fundamental parameter errors through Monte-Carlo technique we perform the fit for each data set fifty times, each time re-sampling from the original data set with replacement, to perform a bootstrap procedure. The isochrones are also re-generated in each run from the adopted IMF and binary fraction of 50$\%$ as described in Paper I and Paper II, and in \citet{Monteiro2011}. The final uncertainties in each fundamental parameter are then adopted to be the standard deviation of the results of all runs.

In the case of the UBVRI photometry, we follow the traditional route of first estimating the reddening through the $(U-B)versus(B-V)$ diagram. In the following steps, where all parameters are allowed to vary, the procedure constrains the reddening parameter $E(B-V)$ to be within $\pm 10\%$ of this initial estimate.

In this work we include the binary fraction as a free parameter for the fits using Gaia data. The fits also adopt the same reddening law from \cite{CCM89} as used in our previous works. The parameter $E(B-V)$ and $R_V$ are determined directly from the fits since they are free parameters. The $A(\lambda)$ for each Gaia filter, which is related to $E(B-V)$ and $R_V$ is obtained from the relations of \cite{CCM89}.

\begin{table}
\caption{Cluster structural parameters used in the isochrone fit. The first two columns (after the cluster identification) give the central coordinates used in J2000.0. The following columns
give the characteristic cluster radius ($R_{c}$) in arcminutes, which is calculated based
on the distribution of distances of each sample star to the cluster center and the width (${SIG\_PIX}$) of the Gaussian kernel used in the star density distribution determination. The clusters marked
with an asterisk had their boundaries determined from the density distribution ($\rho$) given in the last column and not the estimated characteristic radius listed.}
\label{tab:ceparameters}
\begin{tabular}{lccccc}
\hline
\noalign{\smallskip}
Cluster & RA  &   DEC  & $R_{c}$             &   ${SIG\_PIX}$  &   $\rho$    \\ 
\hline
\noalign{\smallskip}
Collinder~307  & 248.833       &  -51.0000    &   2.0      & 0.6   & 13.8    \\
Dias~4~*       & 205.788       &  -63.0499    &   2.5      & 0.6   & 10.4    \\
Dias~6~*       & 277.612       &  -12.3291    &   1.7      & 0.6   & 12.0    \\
IC~4651        & 261.208       &  -49.9390    &   5.0      & 0.6   & 23.4    \\
NGC~5138       & 201.733       &  -59.0594    &   3.5      & 0.6   & 14.2    \\
NGC~6087~*     & 244.629       &  -57.9645    &   5.0      & 0.7   & 20.0    \\
NGC~6178~*     & 248.892       &  -45.6447    &   4.0      & 0.6   & 17.6    \\
\hline
\noalign{\smallskip}
\end{tabular}
\end{table}

For the Gaia DR2 data the CE method to fit the theoretical isochrones  uses the same model sets and procedures. However, due to the high precision membership determination possible, there is no need for photometric membership estimation and therefore only the astrometric memberships are used. We have also left the binary fraction in the cluster as free parameter. Given the high precision of the Gaia data we also have considered the photometric errors in the likelihood function used in the fits at the $3\sigma$ level to improve the convergence speed of the code.

The final fit results obtained for each cluster are presented in Table \ref{tab:ceparameters} and in Figures 1 to 7. 
For all clusters but Dias 4 and Dias 6 the final results were obtained using the mean of $P_{ph}$ and $P_\mu$ when using UBVRI data. For Dias 4 and Dias 6 we used $P = P_{ph}$ as discussed in the next section. The memberships used for the Gaia data were obtained as described in Sec. 4.1.

\section{The color-magnitude diagrams}

The results as well the membership analysis obtained for each individual cluster are discussed in more detail below. All data and the memberships used in the isochrone fits presented in this work are available electronically in our survey website\footnote{\url{https://wilton.unifei.edu.br/OPDSurvey.html}}.

\begin{figure*}
\centering
\includegraphics[scale=0.44]{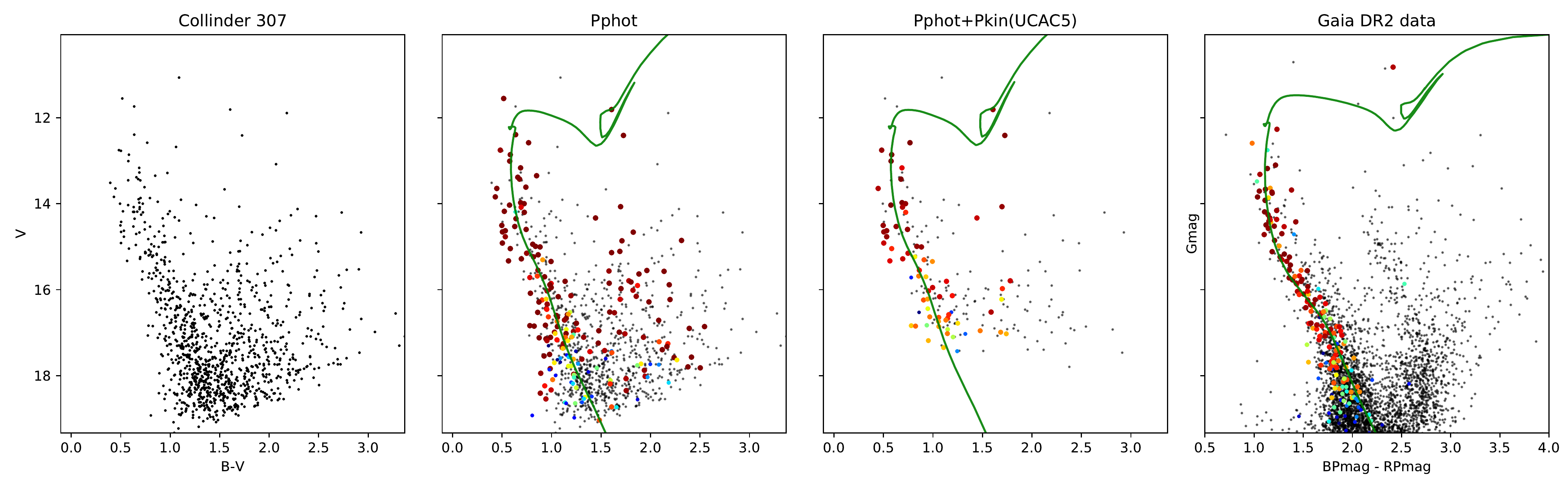}
\caption{Color-magnitude diagram of the Collinder 307. From the left to the right, panel 1 presents the observed stars with no membership determination. Panel 2 presents the same stars but with membership estimated by the $P_{ph}$. Panel 3 presents the stars identified in the UCAC5 catalog which received membership considering $P_{ph}$ and $P_{\mu}$. Panel 4 presents only data from the Gaia DR2 catalog. The memberships were determined from astrometric data (parallax and proper motion) considering the errors and correlations following the Equation \ref{eq:multinormal}.
See the text for details. The data and a complete set of plots, including the color-color diagram, are available at the web-site of the
project {\url{https://wilton.unifei.edu.br/OPDSURVEY/Collinder307-13jul07.html}}}
\label{fig:Collinder307}
\end{figure*}  

\subsection{Collinder 307}

Figure \ref{fig:Collinder307} shows the CMDs of the observed stars with a clear MS extended from $V \approx 12 $ to 18. However, it is not clear in panel 1 if the TO is at $V \approx 12 $ or $V \approx 14 $. The MS is wide due the field contamination as well as due to the increase in photometric errors down to $V \approx 15 $. 

In the same Figure \ref{fig:Collinder307} we show the CMDs comparing the results obtained using  $P_{ph}$ and the mean of $P_{ph}$ and $P_{\mu}$ to determine the final membership.  
Note that using the membership by $P_{ph}$ the possible TO in $V \approx 14 $ is discarded since the stars with V between 14 and 15 and $(B-V)$ between 1.0 and 1.3 received low $P_{ph}$. In the middle panel $P_{ph}$+$P_{\mu}$(UCAC5) the members determined by $P_{ph}$ and $P_{\mu}$ contribute to decontaminate the CMD and to define this feature of the cluster. The final results of the isochrone fit by the CE presented in the middle panels $P_{ph}$ and $P_{ph}$+$P_{\mu}$(UCAC5) do not show statistically significant differences, but we obtained smaller uncertainties using final P as the mean of $P_{ph}$ and $P_{\mu}$. 

Our team published the first CCD UBVRI data for the stars in the region of Collinder 307 in \citet{Monteiro2017}. The estimated parameters, guided by $P_{ph}$, reproduced in Table \ref{tab:results} put the cluster at about 1100 pc and age of logt = 8.60 yr. 
In this work the cluster was used as control object since we include extra steps in constraining the membership probabilities.

As presented in Table \ref{tab:results}, the distance of 1308 pc and logt = 8.65 yr obtained in this work agree very well with the one previously obtained in \citet{Monteiro2017} and the parameters determined by \citet{Carraro2006}. Note that our E(B-V) is estimated using all of the UBVRI photometry after an initial estimate by the color-color (U-B)versus(B-V) diagram while \citet{Carraro2006} used V and I filters. 

In the Figure \ref{fig:Collinder307} we also present the results obtained from the Gaia DR2 data. The memberships determined from the stellar parallaxes and proper motions by the Eq. \ref{eq:multinormal} given in the Sec. 4 show a clear MS and TO at $Gmag \approx 13 $. The lower scattering of the data in the MS is visible due to the low uncertainty of the Gaia photometric data and high definition in the membership determination. Note that none of the giant field star ($BPmag-RPmag \geq 2.5$) was selected as high probability member.
The values of the parameters obtained in our best isochrone fit are presented in Table  \ref{tab:results}. The results from UBVRI and Gaia data agree well within the errors, indicating that Collinder 307 is a intermediate age cluster with a binary fraction of $(40\pm10)$\% situated inside the Carina spiral arm.

\subsection{Dias 4}

Figure \ref{fig:Dias4} shows the CMD of the observed stars in a region of 10 arcmin centered in the coordinates of the cluster Dias 4, published in DAML02.
It exhibits a wide MS and no obvious TO location.
 
\begin{figure*}
\centering
\includegraphics[scale=0.44]{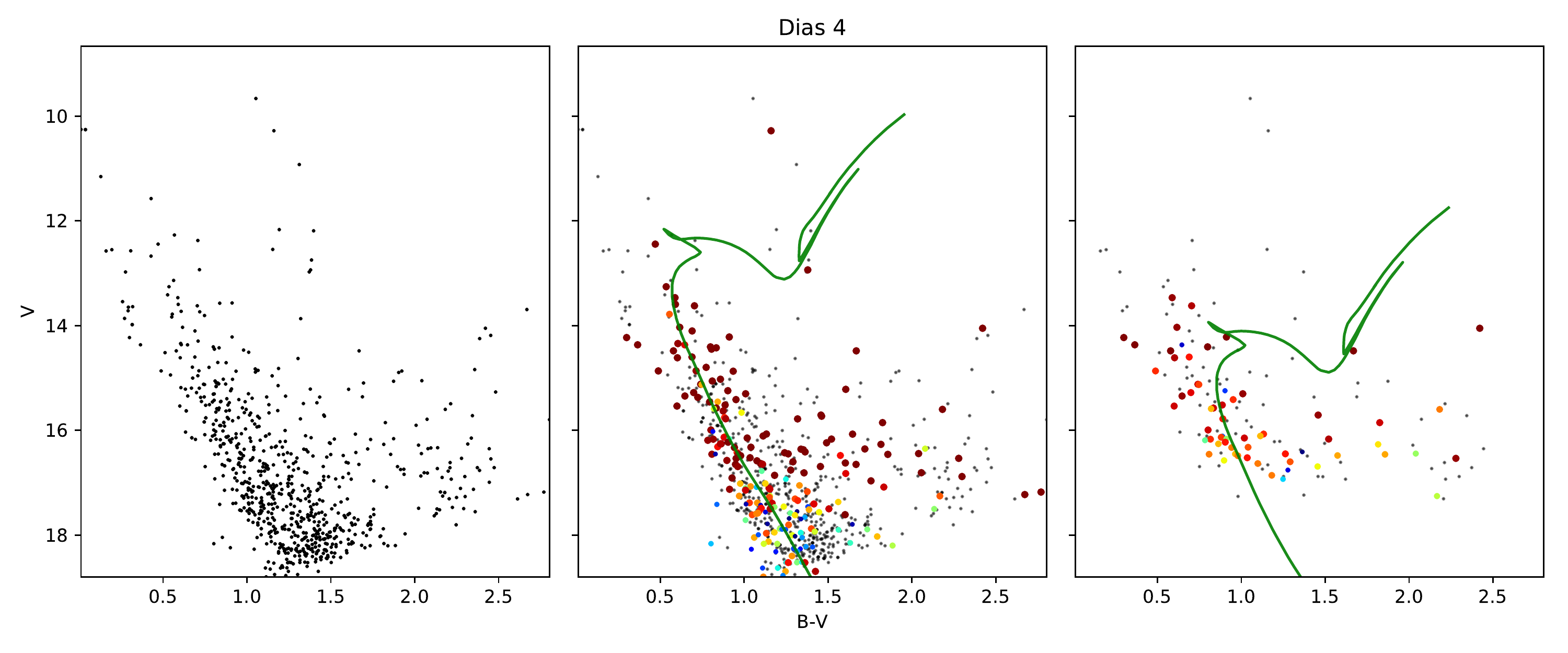}
\caption{Same as Figure 3, but for Dias 4. The data and a set of plots, including the color-color diagram, are available at the web-site of the
project {\url{https://wilton.unifei.edu.br/OPDSURVEY/Dias4-13jul05.html}} }
\label{fig:Dias4}
\end{figure*}  

The results of membership estimation using the photometric data $P_{ph}$ performed by our code is given in the middle panel of the Figure \ref{fig:Dias4}. It  eliminates most of the bright stars since they have positions outside of the adopted iso-density of $10.4~stars/arcmin^2$ for this cluster.
However, one bright star, TYC 9008-3894-1, remains with high $P_{ph}$ membership probability. On the other hand, allowing this bright star as member in the final data to be fitted would imply a large gap in the MS between V magnitudes of 11 and 12. Since such gaps are very unlikely given what is known of stellar evolution, we set $P_{ph}=0$ for it. 
This interpretation is confirmed by the members estimated by the mean of $P_{ph}$ and $P_{\mu}$ presented in the right panel of the Figure \ref{fig:Dias4}.
In this particular case, the membership estimated using $P_{ph}$ and $P_{\mu}$ is limited because a small part of the MS is sampled. Also, for magnitudes above $V \approx 14$ there is high field star contamination which, coupled with the fact that it is a region of greater photometric and kinematic errors, leads to uncertain membership probabilities. For Dias 4 the poor membership estimated implies in results with greater uncertainties.  
%tem que colocar um comentario enfatizado que pelas caracateisitcas do cluster os membership obtidos por Pphot e Pmu nao sao satisfatorios.

The parameters obtained using CCD UBVRI photometry agree with the fit weighted by the $P_{ph}$ and the mean of $P_{ph}$ and $P_{\mu}$, but we opted for the values given by the $P_{ph}$ due to better memberships estimated and smaller final errors in the distance and age. The intermediate age is in agreement with K13 but our results indicate that the cluster should be much closer at about 1300 pc.

The lack of definition presented in the UBVRI photometry was resolved when we inspected the Gaia DR2 data. However in the process of studying this supposed cluster we have discovered that the field around this position has not one but two clusters. The first indication was given when we extracted the same field from the Gaia DR2 release as observed in the UBVRI. Our code determined the membership as described previously and the isochrone was fit but the result we obtained gave a much younger age. This prompted us to investigate the field in more detail, which we did using the software Topcat \citep{topcat}. Upon analyzing a larger 20 arcmin field around the Dias 4 central position we found that indeed there were two density clumps in the ($\mu_{\alpha}cos{\delta}$, $\mu_{\delta}$, $\varpi$) data space as can be seen in Fig. \ref{fig:topcat3D}.

\begin{figure}
\centering
\includegraphics[scale=0.35]{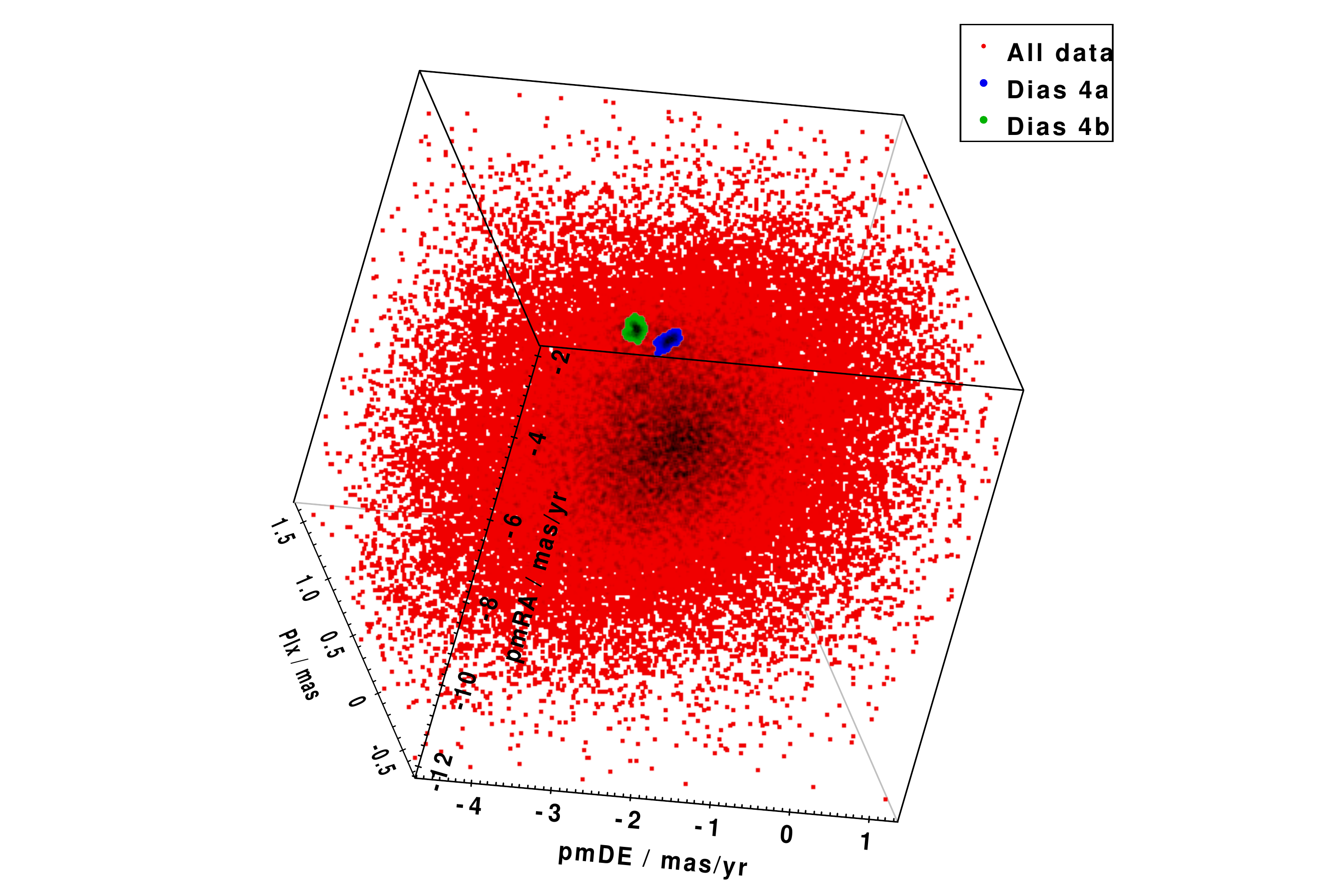}
\caption{The ($\mu_{\alpha}cos{\delta}$, $\mu_{\delta}$, $\varpi$) data in a field of 20 arcmin radius around the center of the Dias 4 coordinates showing the two distinct density enhancements detected.}
\label{fig:topcat3D}
\end{figure}

Using Topcat we selected the stars belonging to these two clumps and inspected their distribution in the field. As shown in Fig. \ref{fig:topcat3Dposition}, there are two clear star density enhancements in position as well as the proper motion and parallax mentioned previously. 

\begin{figure}
\centering
\includegraphics[width=\columnwidth]{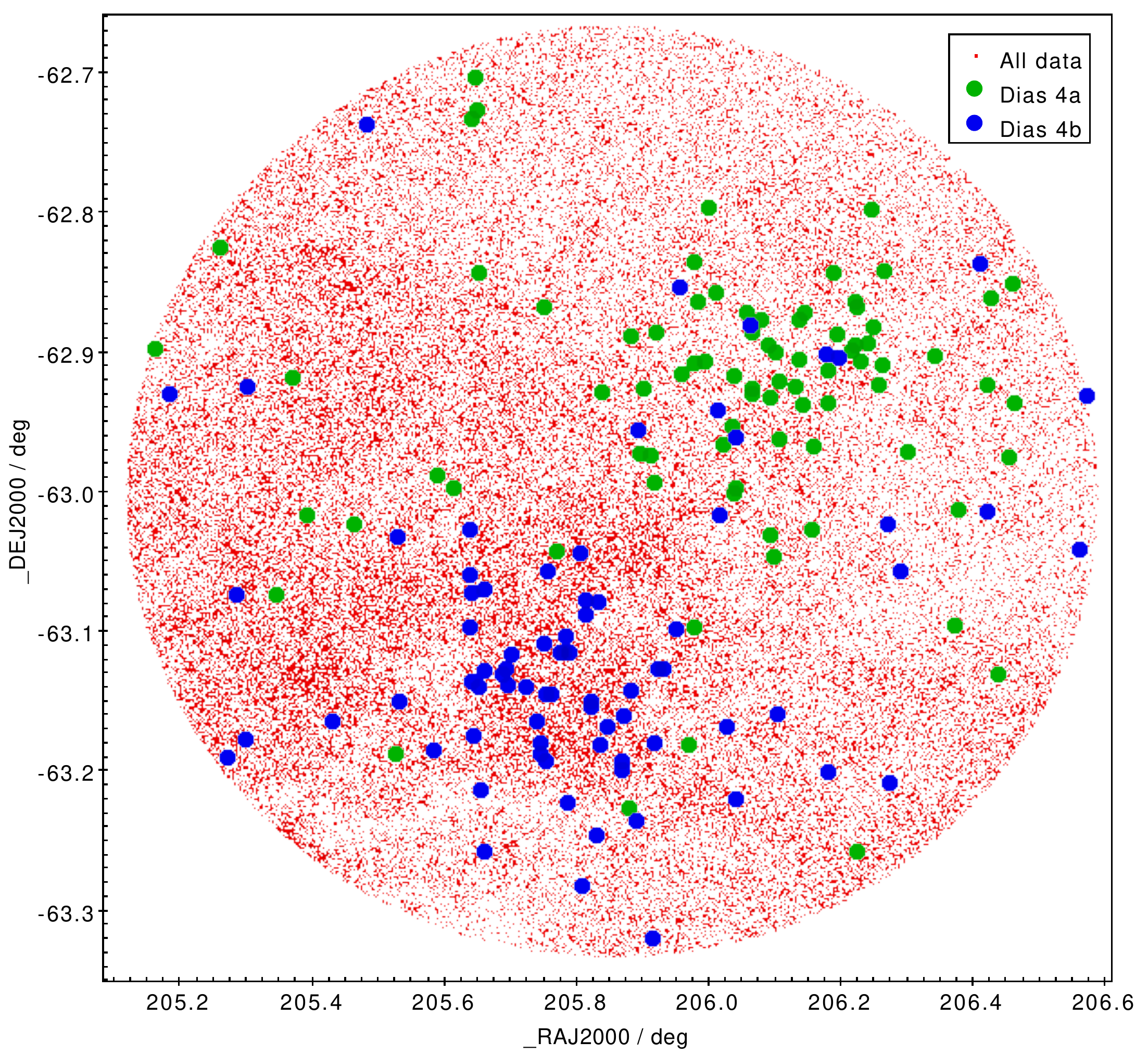}
\caption{The position distribution of the two  ($\mu_{\alpha}cos{\delta}$, $\mu_{\delta}$, $\varpi$) data density peaks showing the resulting two distinct spatial density enhancements.}
\label{fig:topcat3Dposition}
\end{figure}

With the two distributions in space defined we obtained their center coordinates and estimated radius and used those as input to obtain two distinct fields from the Gaia DR2 data. These fields with slightly different centers from the original one of Dias 4 were analyzed and fitted by our isochrone fit code. In Fig. \ref{fig:dias4ab} and Tab. \ref{tab:results} we show the results of the isochrone fits for the two distinct regions. The fits show very clearly two distinct clusters with clear main sequences of distinct ages, one older than the other. The two clusters are at about the same distance, roughly same metallicity and reddening.

The inspection in the literature including the clusters recently discovered by 
\citet{Cantat-Gaudin2018arXiv180508726C} and \citet{Castro-Ginard2018arXiv180503045C} using Gaia DR2 catalog revealed that Dias 4a and Dias 4b were not previously detected star clusters.

\begin{figure*}
\centering
\includegraphics[scale=0.6]{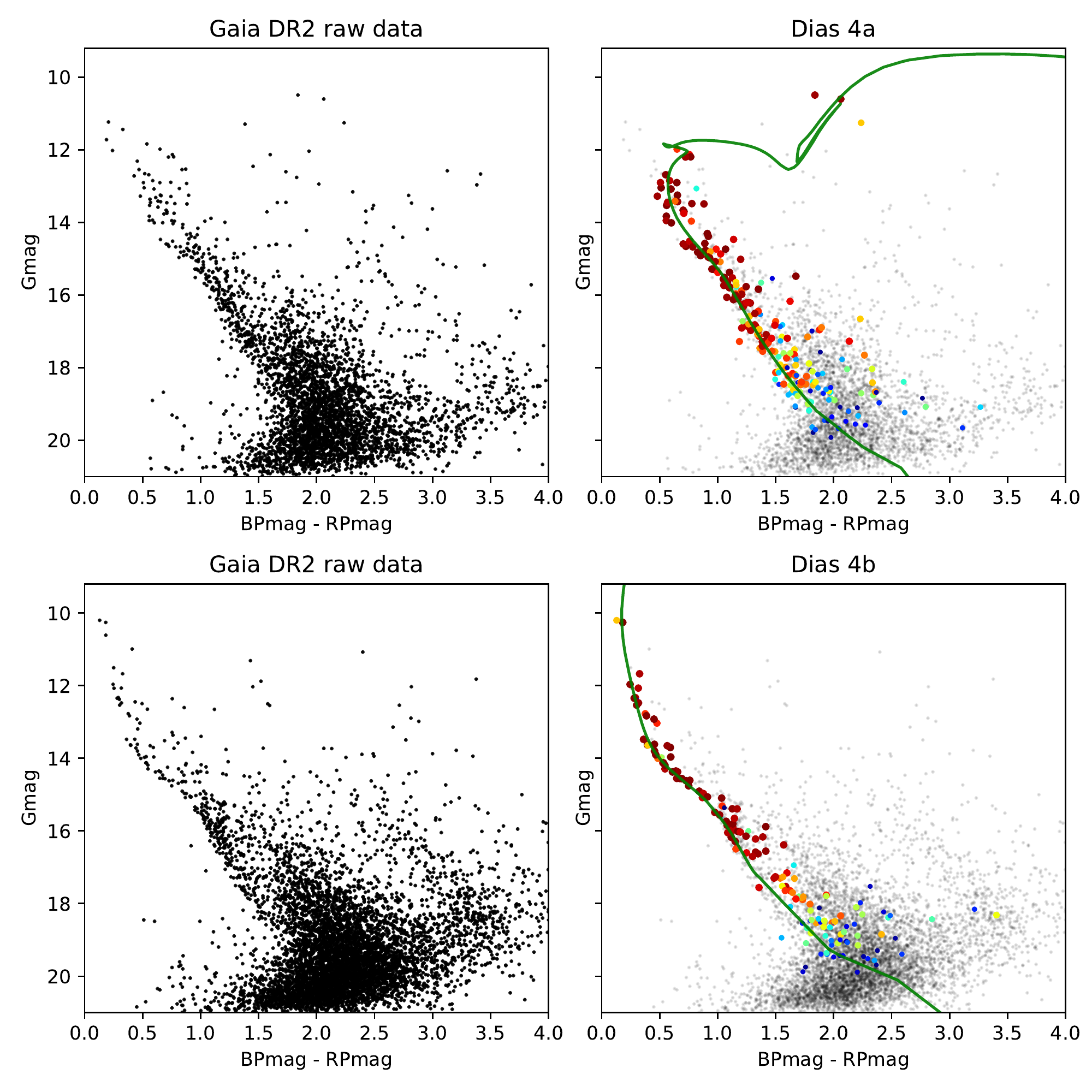}
\caption{Isochrone fit results to the Gaia DR2 data for the two new Clusters Dias 4a and 4b.}
\label{fig:dias4ab}
\end{figure*}

\subsection{Dias 6}

Figure \ref{fig:Dias6} presents in the left panel the CMD of the observed stars with a clear MS ranging from $V \approx 14$ to 18, the TO at $V \approx 14.5$ and a clump of stars (red giants) at $V \approx 15$, $(B-V) \approx 2.0$. 

The scattering in the CMD probably occurs due to both the field stars contamination and the increase in photometric errors for $V \ge 15$.

In the $P_{ph}$ and $P_{ph}$+$P_{\mu}$(UCAC5) panels of the Figure \ref{fig:Dias6} we present the same data weighted by the $P_{ph}$ and the mean of $P_{ph}$ and $P_{\mu}$. 
As commented in Table \ref{tab:ceparameters} we used iso-density of $12.0~stars/arcmin^2$ to select the most probable member stars. However some bright blue stars received high $P_{ph}$ due to their position close to the determined cluster center. 
% pelo que vejo na figura a frase abaixo nao esta correta! E melhor retira-la. Avisamos ao referee?
%To avoid biasing the fit to younger ages due to these stars, since the TO is clearly located at $V \approx 14.5$, we set $P = 0$ for the stars with $V \leq 14$.  

Weighing the data by $P_{ph}$ and $P_{\mu}$ makes the MS and the location of the TO of the cluster in the CMD less clear, mainly due to photometric and proper motions errors for stars with $V \geq 15$. 
The best isochrone fit performed with these data agrees within $1 \sigma $ with the results obtained using $P_{ph}$ but it is more uncertain due the factors mentioned above. As final results given in Table \ref{tab:results} we opted for the fit weighted by the $P_{ph}$ since it maintains the cluster feature in the CMD more reliably and because it presents smaller errors. 

In the right panel of the Figure \ref{fig:Dias6} we present the Gaia DR2 data with the membership determined from the parallaxes and proper motions of the stars as detailed in the Sec. 4. Note that the Gaia data and the membership determination better define the giants stars which impose greater constraint to age determination by isochrone fit.
The final result obtained by the CE method is given in the Tab \ref{tab:results}. It is very interesting the agreement between the results obtained with the UBVRI and Gaia data which we also attribute in part to the performance of the non-subjective CE method used in the isochrone fit.

We point out the value of E(B-V) obtained by the UBVRI data is probably underestimated due to the contamination of bright blue stars which received high $P_{ph}$. However, the results confirm the values obtained by \citet{Dias2012} using preliminary UBV data, and shows that Dias 6 is an intermediate age cluster with a binary fraction of ($58\pm8$)\% located at about 2700 pc, unlike the values estimated in the literature based on 2MASS data only.

\begin{figure*}
\centering
\includegraphics[scale=0.44]{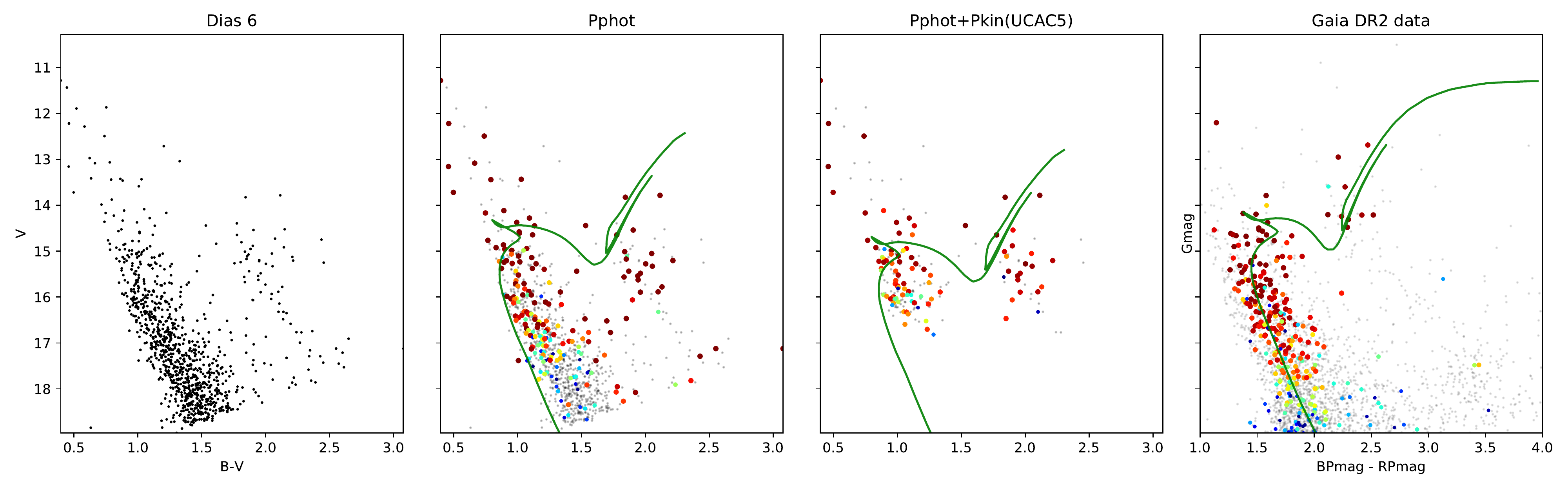}
   \caption{Same as Figure 3, but for Dias 6. The data and a complete set of plots, including the color-color diagram, are available at the web-site of the
project {\url{https://wilton.unifei.edu.br/OPDSURVEY/Dias6-13jul06.html}}}
\label{fig:Dias6}%
\end{figure*}

\subsection{IC 4651}

The CMD presented in Figure \ref{fig:IC4651} is easy to interpret because it displays a clear MS between $V \approx 12$ and 15 and TO at $V \approx 12$. The field contamination is most evident at $V \ge 16$. 

In the middle panels of the Figure \ref{fig:IC4651} we present the CMD with the best isochrone fit obtained  by the photometric data weighted by $P_{ph}$ and by the mean of $P_{ph}$ and $P_{\mu}$ respectively. In this analysis we opted to use only the stars with $V \le 16.0 $ and $(B-V) \le 1.15 $ to exclude field stars that received high P values in our method. In this way we avoid greater bias in the distance estimates due to the contribution of distant field stars at $V \ge 16.0 $ and in the age due giants at different distances and with different color excesses. There is no statistical differences between the results obtained using $P_{ph}$ and $P_{ph}$ and $P_{\mu}$ weighing the isochrone fit but we opted to use the results with smaller errors. 

As the cluster is well studied it was possible to compare our $P_{ph}$ and $P_{\mu}$ membership with those determined from radial velocities for 6 stars in common with \citet{Mermilliod1995} and 78 stars in common with \citet{Meibom2002} with V mags ranging between 10 and 14.
All membership probabilities we estimate agree with those from \citet{Mermilliod1995}. In the comparison with \citet{Meibom2002}, only 7 member stars ($P \ge 51\%$) were misclassified by our method.

The best isochrone fit obtained by our code presented in Table \ref{tab:results} and Figure \ref{fig:IC4651} was based on 175 stars with $P \ge 51\%$ determined by $P_{ph}$ and $P_{\mu}$. Our values show good agreement with those published, in particular with more recent works that estimate larger cluster distances of about 1000 pc and younger age.

The analysis performed with the UBVRI data is supported by the results obtained with the Gaia DR2 data for which the results are presented in the right panel of the Figure \ref{fig:IC4651} and in Table \ref{tab:results}. The selected member stars are sufficiently accurate to distinguish a very well defined main sequence with a visible binary sequence spread with a fraction estimated as ($50\pm8$\%) by our method .

Our final value of the metallicity (Z) parameter obtained from the best isochrone fit is transformed to $[Fe/H] = (0.11 \pm 0.13)$ dex (from UBVRI data) and $[Fe/H] = (0.11 \pm 0.12$) dex (from Gaia data), adopting the same conversion as considered in the Padova database of stellar evolutionary tracks and isochrones: $[Fe/H]=logZ/Z\odot$ with $Z\odot=0.0152$. This is in agreement with the value $[Fe/H] = (0.11 \pm 0.01)$ dex published by \citet{Carretta2004} determined from photometric and spectrometric data for five evolved stars of the cluster. Considering the uncertainty, our estimate agrees with the values published in the literature which range from 0.077 to 0.23 dex as compiled by \citet{Mikolaitis2011MNRAS.413.2199M}.

%It is interesting to note about the [Fe/H] estimated by the our isochrone fit although its not the main subject of this study. 
%As compiled by \citet{Mikolaitis2011MNRAS.413.2199M} $[Fe/H] = (0.11 \pm 0.01)$ dex was determined from photometric and spectrometric data for five evolved stars of the cluster \citep{Carretta2004}. 
%Our final value of the metallicity (Z) parameter obtained from the best isochrone fit is transformed to $[Fe/H] = (0.11 \pm 0.13)$ dex (from UBVRI data) and $[Fe/H] = (0.11 \pm 0.12$) dex (from Gaia data), adopting the same conversion as considered in the Padova database of stellar evolutionary tracks and isochrones: $[Fe/H]=logZ/Z\odot$ with $Z\odot=0.0152$. This is in excellent agreement with the value published by \citet{Carretta2004} and considering the uncertainty, our estimate agrees with the values published in the literature which range from 0.077 to 0.23 dex.

\begin{figure*}
\centering
\includegraphics[scale=0.44]{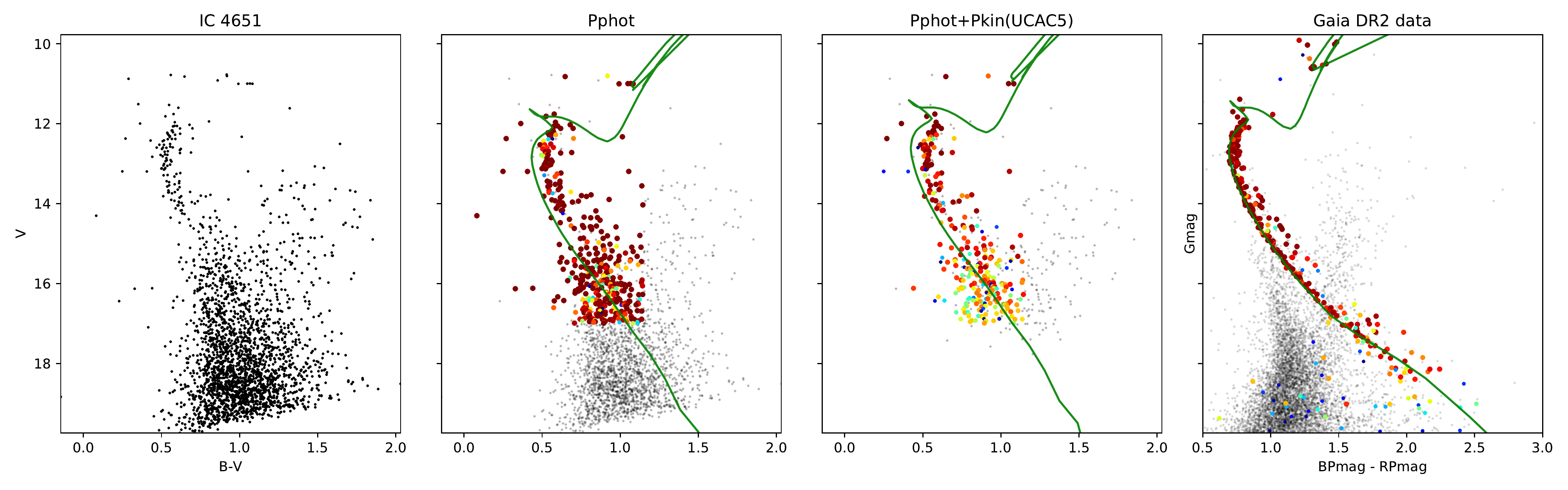}
   \caption{Same as Figure 3, but for IC 4651. The data and a complete set of plots, including the color-color diagram, are available at the web-site of the
project {\url{https://wilton.unifei.edu.br/OPDSURVEY/IC4651-13jul05.html}}}
\label{fig:IC4651}%
\end{figure*}

\subsection{NGC 5138}

In Figure \ref{fig:NGC5138} we present the CMD of the observed stars where the feature of the cluster is clear, with the MS ranging from 11.5 to 17.0 mag. There is a wider spread of the MS between V = 14 and 16, likely due to the higher photometric and proper motions errors and also due to the presence of field stars, which received high membership by our method.   

\begin{figure*}
\centering
\includegraphics[scale=0.44]{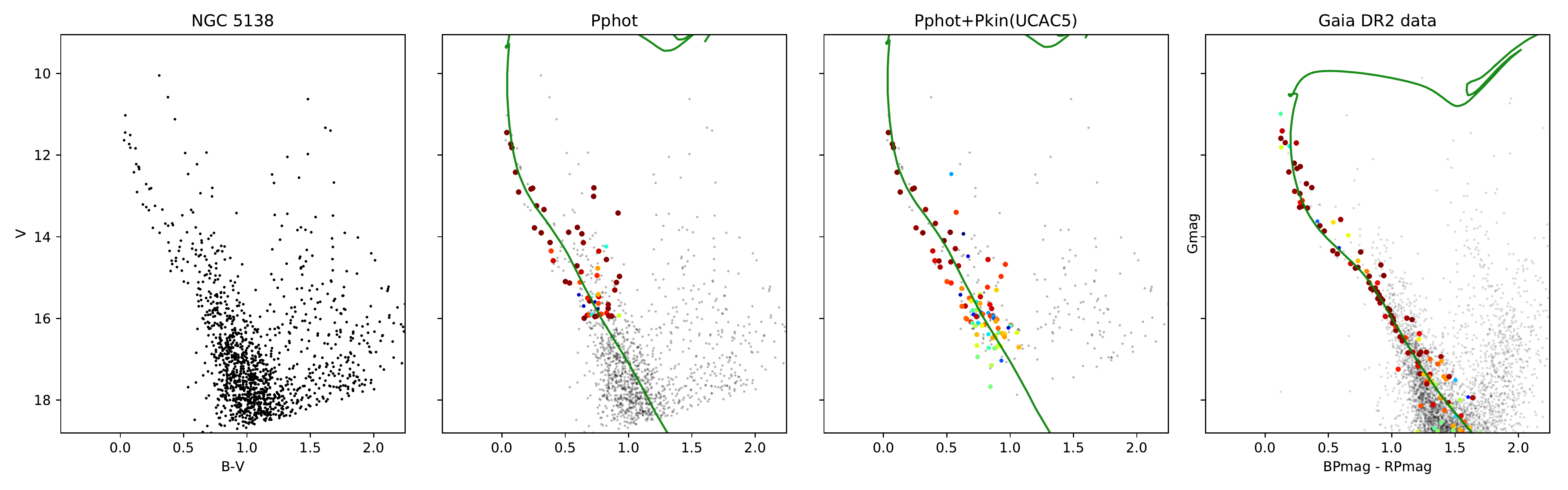}
   \caption{Same as Figure 3, but for NGC 5138. The data and a complete set of plots, including the color-color diagram, are available at the web-site of the
project {\url{https://wilton.unifei.edu.br/OPDSURVEY/NGC5138-13jul05.html}}}
\label{fig:NGC5138}%
\end{figure*}

In the middle panels of the Figure \ref{fig:NGC5138} we present the data weighted by $P_{ph}$ and the mean of $P_{ph}$ and $P_{\mu}$. 
In the analysis using $P_{ph}$ we do not use stars with $(B-V)\geq 1.15$ and $V \ge 16.0$ which are clearly regions with high field star contamination and our method failed to attribute low membership.
However the method attributed low membership to the giant stars ($B-V\geq 1.15$) at different distances and reddening taking into account the membership estimated by the mean of $P_{ph}$ and $P_{\mu}$. 
 Interestingly, the membership estimated by $P_{ph}$ and $P_{\mu}$ excludes a possible interpretation of the TO in the region of $V \approx 12$ and $(B-V) \approx 0.6$. 

Basically, the analysis using memberships estimated by $P_{ph}$ and the mean of the $P_{ph}$ and $P_{\mu}$ lead to results that are very similar but we choose as final results the best isochrone fit using $P_{ph}$ and $P_{\mu}$ (see Tab. \ref{tab:results}), since it was more efficient in excluding the field stars and providing results with smaller errors. 

In the right panel of the Figure \ref{fig:NGC5138} we give the Gaia DR2 data with the membership estimated and the best isochrone fit determined by our method. The members selected define a clear MS of the cluster eliminating the stars with $(B-V)\geq 1.15$ and $V \ge 16.0$ which our method failed to attribute low membership using UBVRI observed data. It is a good example to illustrate the quality of the Gaia data for the membership determination which allows a reliable isochrone fit and estimated binary fraction of ($40\pm7$)\% for the cluster.

The results of the isochrone fit presented in Table \ref{tab:results} obtained from UBVRI and Gaia data are consistent but the difference in the distances is out of $1 \sigma$. The difference  is likely due the wider spread of the MS between V = 14 and 16 in UBVRI data, which could be field stars misclassified as members.

%agrees very well with those obtained from UBVRI observed data. Both are in disagreement with closer distance and younger age published previously in the literature. 

\subsection{NGC 6087}

The CMD of the observed stars presented in the left panel of the Figure \ref{fig:NGC6087}  shows a clear open cluster feature with the MS ranging from $V \approx 8$ to 16 and the Cepheid S Nor which is the brightest observed star.

In our observations, the S Nor was detected in the UBV filters but not in the R and I filters due to saturation. Even so, the star received high $P_{ph}$ value in the analysis using only the photometric data. Surprisingly we found $P_{\mu}= 0\%$ for this star using the proper motion ($\mu_{\alpha}\cos\delta = 2.0 \pm 1.8$ and $\mu_{\delta} = -46.0 \pm 1.9$ in $mas~yr^{-1}$) from the UCAC5 catalog,  which is in disagreement with the Gaia DR2 value ($\mu_{\alpha}\cos\delta = -1.415 \pm 0.066$ and $\mu_{\delta} = -2.287 \pm 0.056$ in $mas~yr^{-1}$).

As different data indicate that S Nor is a real cluster member of the cluster NGC 6087, in our analysis using UBVRI data we set $P_{kin}= 100\%$ for this star, opting to use Gaia's proper motion.

In the analysis considering $P_{ph}$ and the mean of $P_{ph}$ and $P_{\mu}$ we opted to not use stars with $ (B-V) \geq 1.15 $ and $ V \geq 16 $ to eliminate field stars that our method failed to filter out in the membership determination. We notice that approximately the same set of stars was considered as member using $P_{ph}$ and $P_{ph}$ and $P_{\mu}$ and the final values obtained for the fundamental parameters do not differ significantly. 
We choose the values obtained by $P_{ph}$ and $P_{\mu}$ as given in Table \ref{tab:results} and in the Figure \ref{fig:NGC6087} as final results which agree, within the errors, with the published ones.

In the right panel of Figure \ref{fig:NGC6087} we present the Gaia DR2 data over-plotted with the best isochronal fit obtained by our code. The MS of the cluster is evident with the Cepheid S Nor having high membership probability. The parameters estimated by the isochrone fit agrees very well with those obtained using the UBVRI data indicating the cluster NGC 6087 intermediate age one closer than 1kpc from the Sun.

\begin{table}
\caption[]{Comparison of the distances of NGC 6087 obtained from different methods and data. 
In the column 1 is given the code referring to the method used. Column 2 shows the origin of the data used. 
In column 3 are given the distance and the uncertainty. 
In the columns 4 and 5 are given the lower and upper bounds of the about $68\%$ confidence interval. In the last column are given the objects used in the distance estimation. We consider member stars those with $P \geq 51\%$.
The distances obtained from parallaxes were estimated by the procedure published by \citet{Bailer-Jones2018AJ....156...58B} using a distance prior that varies smoothly as a function of Galactic longitude and latitude according to a Galaxy chemo-dynamical model, including extinction. The standard errors provided in the distance from parallaxes were estimated considering the same prior as symmetric the calculated the 5th and 95th percentile confidence intervals.
The uncertainties were calculated using 
$\sigma = r_{95} - r_{5}/(2x1.645)$, which is equivalent to 1$\sigma$ Gaussian uncertainty.}
\label{tab:compdistancesNGC6087}
\begin{center}
\footnotesize
\begin{tabular}{lccccl}
\hline
\hline
method & data used &   distance (pc)&  $r_{5}($\%$)$    &  $r_{95}($\%$)$    & object \\
\hline
2      &  Gaia     &   $913 \pm 64$ &        &         &    members       \\
2      &  UBVRI    &   $1009\pm 97$ &        &         &    members      \\
1      &  Gaia     &   $986 \pm 47$ &   918  & 1074    &    members    \\
1      &  Gaia     &   $944 \pm 42$ &   882  & 1022    &    Snor      \\
1*     &  Gaia     &   $919 \pm 22$ &   884  &  956    &    Snor     \\
1*     &  Gaia     &  $1015 \pm 50$ &   943  & 1109    &    members   \\
3      &  -        &   $814 \pm 23$ &        &         &    Snor       \\
%1*     &  Gaia     &  $972  \pm 46$ &   906  & 1058    &    Snor \\
\hline
\end{tabular} 
\end{center}
\begin{flushleft}
\tiny
method 1 = distance estimated from parallax; 
method 2 = isochrone fit; 
method 3 = value obtained from period-luminosity relation from  \citet{Groenewegen2013}. The methods with flag (*) were corrected from zero point parallax of $-0.029$ mas.
\end{flushleft}
\end{table}

In Table \ref{tab:compdistancesNGC6087} we compile the values of distances related to the cluster NGC6087 obtained through different methods. We show values obtained using isochrone fits and average parallax of members stars with memberships probabilities of $P \geq 51\%$ and  for the S Nor Cepheid using period-luminosity relation.

Even though the results depend on the prior, the values are consistent and in general agree within 1 $\sigma$ level. These results provide an independent check of the distance estimated by our isochrone fit indicating that our methods yields reliable results.

%Finally, our estimated value of the distance for the cluster using the Gaia DR2 data ($913 \pm 64$ pc) agrees with those obtained for S Nor, determined from the pulsation period of the Cepheid ($814 \pm 23$) pc by \citep{Groenewegen2013}. Note that the result based on the Gaia DR2 parallaxes provided by the method published by \citet{Astraatmadja2016} is ($944\pm 63$) pc considering  the exponentially decreasing space density prior and no systematic zero-point correction. These results provide an independent check of the distance estimated by our isochrone fit indicating that our method yields reliable results. 

 \begin{figure*}
 \centering
 \includegraphics[scale=0.44]{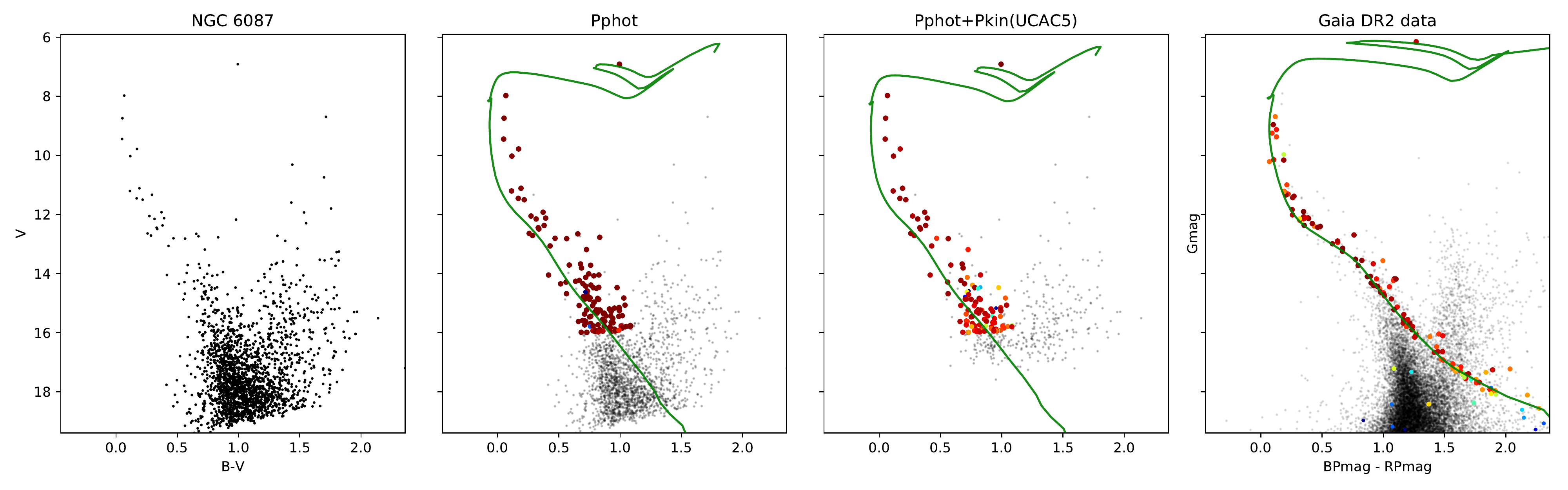}
 \caption{Same as Figure 3, but for NGC 6087. The data and a complete set of plots, including the color-color diagram, are available at the web-site of the
project {\url{https://wilton.unifei.edu.br/OPDSURVEY/NGC6087-13jul06.html}}}
 \label{fig:NGC6087}
 \end{figure*}

\subsection{NGC 6178}

We present for the first time UBVRI CCD data of the stars in the field of NGC 6178. The CMD of the observed stars presented in the left panel of the Figure \ref{fig:NGC6178} shows a clear MS from $V \approx 8$ to 16 with distant field stars contamination at $ V \geq 15.5$. 

In our analysis we opted to artificially set the stars with $ (B-V) \geq 1.5$  and $ V \geq 16$ as field stars to avoid biasing the isochrone fits to greater distances. Using $P_{ph}$ and the mean of $P_{ph}$ and $P_{\mu}$ to weight the data, our method obtains statistically similar fundamental parameters. In the middle panels of the Figure \ref{fig:NGC6178} are presented the data weighted by $P_{ph}$ and by the mean of $P_{ph}$ and $P_{\mu}$. 
In Table \ref{tab:results} we give the final values determined using $P_{ph}$ and $P_{\mu}$ since it leads to smaller errors in the parameters estimated.

The results obtained using the Gaia data are presented in the right panel of the Figure \ref{fig:NGC6178}. The membership provides a clearly defined MS range from $Gmag \approx 8$ to 19. 

The parameters determined by our method of isochrone fitting agree with those obtained using the same method applied to UBVRI data. The parameter E(B-V) shows a larger difference but still within 2$\sigma$.

Despite the fact that our results give an age in agreement with the values obtained by \citet{Piatti2000} and K13, the distance of the cluster is about 200 pc lower, which is comparable with the value published by \citet{Moffat1973}. We conclude that NGC 6178 is a young open cluster with a binary fraction of ($55\pm10$)\% located at about 820 pc from the Sun.   

\begin{figure*}
\centering
\includegraphics[scale=0.44]{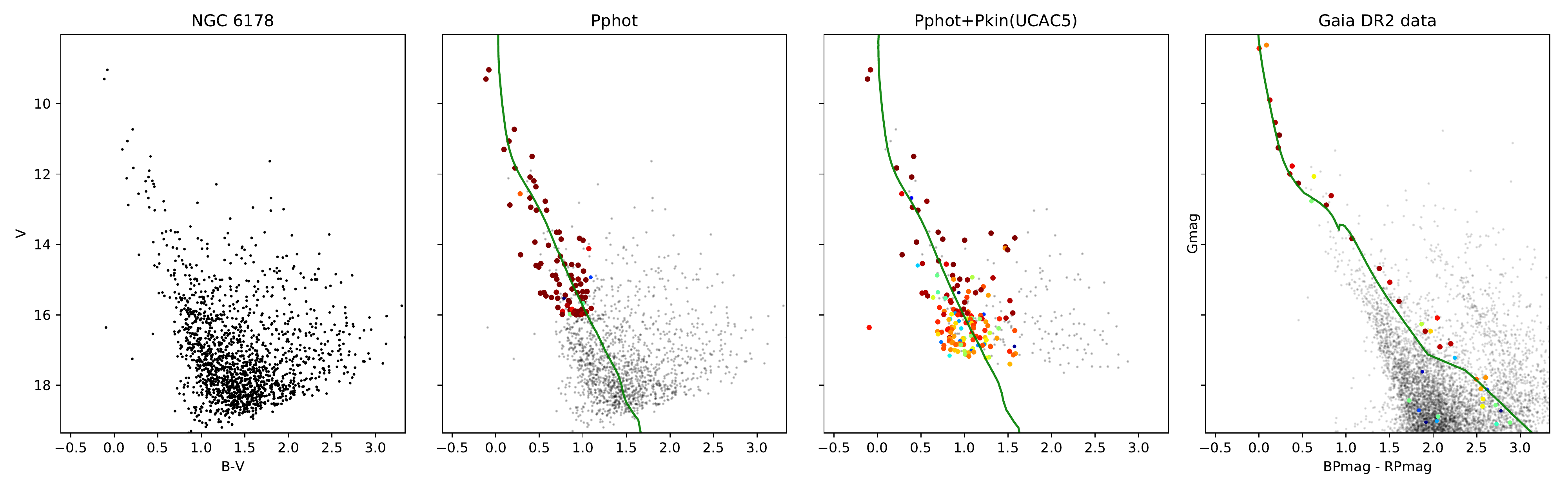}
\caption{Same as Figure 3, but for NGC 6178. The data and a complete set of plots, including the color-color diagram, are available at the web-site of the
project {\url{https://wilton.unifei.edu.br/OPDSURVEY/NGC6178-13jul07.html}}}
\label{fig:NGC6178}%
\end{figure*}

\section{Discussion of the results}

The results presented in the previous section obtained from UBVRI and Gaia DR2 data, show good internal consistency. However, some caveats about the UBVRI results for the clusters studied in this work must be pointed out. 
For the clusters NGC 6087 and IC 4651, the observed field is smaller than the estimated cluster size. This impairs the determination of the $P_{ph}$ since the density map may have inadequate coverage. The errors of the proper motions were used in the determination of $P_{\mu}$ and the errors of the UCAC5 data are considerable for stars with $V \ge 13$. For all investigated clusters, except Dias 4 and Dias 6, the analysis using $P_{\mu}$ was possible because the TO of the cluster is at $V \le 14 $ which allows a good portion of MS to be sampled. For older (and more distant) clusters such as Dias 4 and 6, this does not happen. For Dias 4 and Dias 6 a small part of the MS is sampled in the most contaminated part and of greater errors of the CMD, which leads to greater uncertainties in the final parameters estimated by the isochrone fit using the data weighted by $P_{ph}$ and $P_{\mu}$. Finally, in our photometric method to estimate the memberships, typically bright stars close to the center of the field may receive high $P_{ph}$. Similarly, field stars with proper motion closer to the mean of the cluster may also receive high $P_{\mu}$. For these cases there is no guarantee that the stars will be a real member as happened with the star TYC 8719-650-1 in the field of the cluster NGC 6087. 
All of these ambiguities were solved by the membership determined from astrometric data of Gaia DR2 catalog.

The comparison of the mean proper motion of the clusters determined by UCAC5, UCAC4 and Gaia DR2 data are given in Table \ref{tab:compkin} where we reproduce the results obtained with the catalog UCAC5 of Table \ref{tab:kin} and those obtained with UCAC4 from D14. The mean difference in the mean proper motion from UCAC5 minus Gaia is 0.25 $mas~yr^{-1}$ in $\mu_{\alpha}cos{\delta}$ and 0.13 $mas~yr^{-1}$ in $\mu_{\delta}$. The standard deviations are 0.62 $mas~yr^{-1}$ and 0.45 $mas~yr^{-1}$, in $\mu_{\alpha}cos{\delta}$  and $\mu_{\delta}$, respectively. As expected we found consistent results with the Gaia standard deviation being about ten times lower. 

We also investigated the distances of the clusters obtained by our method of isochrone fitting and those estimated from parallaxes using the Gaia data. 
The comparison is given in Table \ref{tab:compdistances}
where the values estimated from the parallax data were determined using the mean and standard deviation parallaxes given in Table \ref{tab:meanparams}. If we consider this value as the parallax of a single star situated in the same parallax of the cluster we can convert it to distance by the procedure published by \citet{Astraatmadja2016} and \citet{Gaia-DR2-Luri2018}. The distance was estimated using the prior given by the procedure published by \citet{Bailer-Jones2018AJ....156...58B}, which uses a distance scale that varies smoothly as a function of Galactic longitude and latitude according to a Galaxy chemo-dynamical model, including extinction.
The distance estimate is given by the mode and the standard errors provided were estimated considering the calculated 5th and 95th percentile confidence intervals. Using a symmetric distribution approximation we can calculate $\sigma = r_{95} - r_{5}/(2x1.645)$, which is equivalent to 1$\sigma$ Gaussian uncertainty. 

The results agree within the errors indicating that our method is reliable. Interestingly, although the comparison is based on only six points, there is a systematic tendency for the distance from parallax to be greater than the distance from the isochrone fit. The mean difference is of about 28 pc in the sense of distance from parallaxes minus distance from isochrone fit with a standard deviation of 96 pc. 
%This value is equivalent to \textbf{0.028} mas in the parallax. 

The comparison of the results obtained by the isochrone fit using UBVRI and Gaia data show how much the membership from Gaia improves cluster feature detection. The results of mean proper motion are much more precise allowing to go beyond the separation of cluster and field populations, as was the case of Dias 4 where we were able to detect two distinct clusters in the field.

The agreement between the values obtained using UBVRI and Gaia data underlines the quality of our ground based observed data. It is also interesting to point out that in this work the E(B-V) values estimated from the Gaia data agree with the values determined with the photometry using the U filter.  The values of [Fe/H] estimated using UBVRI and Gaia data are also in agreement and the comparison with the literature indicates it can be estimated with uncertainty of about 0.1 dex from the isochrone fit method as shown by \citet{Oliveira2013}.

Finally, we estimated the binary fraction of the clusters to be in the range between 40\% - 60\%  with the Gaia photometric data. This aspect as well as others will be investigated in more detail with a large number of open clusters with different ages in a forthcoming paper.

\begin{table}
\label{tab:compkin}
\caption[]{Comparison of our results for
mean proper motions with those provided
by \citet{Dias2014} and obtained using the Gaia DR2 data.
in the column nstars are given the number of estimated member stars of each cluster. The values are in $mas~yr^{-1}$.}
\begin{center}
\footnotesize
\tiny
\begin{tabular}{lcccccc}
\hline
cluster       &   $\mu_{\alpha}cos{\delta}$&  $\sigma$&  $\mu_{\delta}$&   $\sigma$& nstars &    catalog \\
\hline   
Collinder 307 &  -1.9   &  0.2       & -3.8    &  0.1        & 196    &   UCAC5    \\   
              &  -0.6   &  5.3       & -3.5    &  3.5        & 176    &   UCAC4    \\   
              &  -0.725 &  0.023     & -3.294  &  0.028      & 484    &   Gaia     \\   
\hline
NGC 5138      &  -5.9   &  2.3       & -2.1    &  1.1        & 350    &   UCAC5    \\   
              &  -8.0   &  5.1       & -1.4    &  4.6        & 283    &   UCAC4    \\   
              &  -3.600 &  0.018     & -1.469  &  0.059      & 200    &   Gaia     \\   
\hline
Dias 4        &  -5.7   &  1.1       & -2.2    &  0.1        & 155    &   UCAC5    \\   
              &  -4.3   &  4.8       & -5.3    &  3.5        & 168    &   UCAC4    \\   
Dias4a        &  -4.529 &  0.028     & -1.985  &  0.023      & 1484    &   Gaia     \\   
Dias4b        &  -4.439 &  0.012     & -2.344  &  0.004      & 916    &   Gaia     \\                 
\hline
Dias 6        &  0.8    &  0.7       & -1.8    &  1.3        & 175    &   UCAC5    \\   
              &  -1.0   &  1.9       & -1.6    &  1.8        & 105    &   UCAC4    \\   
              &  0.5076 &  0.021     & -0.580  &  0.023      & 953    &   Gaia    \\                 
\hline
IC4651        &  -2.5   &  0.8       & -5.2    &  0.1        & 552    &   UCAC5    \\   
              &  -1.6   &  3.8       & -2.6    &  4.1        & 480    &   UCAC4    \\   
              &  -2.459 &  0.062     & -5.073  &  0.017      & 814    &   Gaia     \\   
\hline
NGC 6087      &  -2.8   &  2.2       & -4.4    &  2.2        & 972    &   UCAC5    \\   
              &  -5.4   &  4.4       & -4.9    &  4.7        & 620    &   UCAC4    \\   
              &  -1.626 &  0.048     & -2.405  &  0.051      & 1180    &   Gaia     \\   
\hline
NGC 6178      &  -1.2   &  0.9       & -3.0    &  1.5        & 286    &   UCAC5    \\   
              &  -0.8   &  6.3       & -0.8    &  5.0        & 203    &   UCAC4    \\ 
              &   0.594 &  0.075     & -3.382  &  0.019      & 380    &   Gaia    \\
\hline
\end{tabular}
\end{center}
\end{table}

\begin{table}
\caption[]{Comparison of the distances obtained from parallax data given in column two and distances determined with our method of isochrone fit given in the column three both using the Gaia DR2 data. We used the mean and standard deviation parallaxes given in Table \ref{tab:meanparams} converted to distance by the procedure published by \citet{Bailer-Jones2018AJ....156...58B}.
The standard errors provided in the distance from parallaxes were estimated considering a symmetric distribution approximation to the calculated 5th and 95th percentile confidence intervals. The values are given in pc.}
\label{tab:compdistances}
\begin{center}
\footnotesize
\begin{tabular}{lcc}
\hline
cluster       &   distance    &     distance             \\
              &  from parallax   &  from isochrone      \\
\hline
Collinder 307 & $1986 \pm 135 $      & $1819  \pm 497 $      \\   
%Dias 4       & $1896 \pm  84 $      & $1601 \pm  166 $      \\   
Dias 6        & $3115 \pm  568 $     & $2627 \pm  182 $      \\   
IC4651        & $944  \pm   42  $    & $ 924 \pm   52 $      \\   
NGC 5138      & $1939 \pm  197 $     & $1857  \pm  95 $     \\   
NGC 6087      & $986  \pm   47$      & $1009 \pm   97 $     \\ 
NGC 6178      & $901  \pm  41 $      & $ 817 \pm  108 $      \\
   
\hline
\end{tabular}
\end{center}
\end{table}

\section{Conclusions}

We presented fundamental parameters of seven open clusters studied which except for one, had UBVRI photometry obtained for the first time. It is surprising that for some open clusters there were photometric data for only a reduced number of stars, as in the case of the NGC 5138 studied in this work. Although in general photoelectric data are more accurate, the number of observed stars is typically much smaller, which can lead to problems in the distance and age determination due to the poor sampling of the main sequence in the CMD.

The results of the parameters and mean proper motions obtained in this work agree with previous ones from the literature. 

In this work we have also used the Gaia DR2 data to estimate membership from astrometric data which, due to its superior quality, better define the main sequence and turn-off of the clusters in the CMD. It helps to constrain the isochrone to fit the clusters providing more reliable and precise results. From the isochrone fit performed by our global optimization tool we conclude is possible determine the E(B-V) from the Gaia photometric data and estimate values of [Fe/H] with uncertainty of about 0.1 dex.

An interesting point in this work is the consistency between the UBVRI and Gaia results. We also point out the agreement between the distance determined by parallax, period-luminosity relation and the isochrone fit for the Cepheid S Nor, member of the cluster NGC 6087. This external check shows the results obtained by the global optimization tool developed in our previous papers, to fit theoretical isochrones to open cluster photometric data is reliable.

Finally, based on the Gaia data we also report a discovery of two new clusters (Dias 4a and Dias4b) in the extended field near what was originally Dias 4.

\section*{Acknowledgements}
We thank the referee for his/her valuable suggestions which improved the text. The entire project was made possible by large amounts of observing time and travel and other financial support from LNA/MCTI. We thank the staff of the Pico dos Dias Observatory for the valuable support. 
W. S. Dias acknowledges the S\~ao Paulo State Agency
FAPESP (fellowship 2013/01115-6). H. Monteiro would like to thank
FAPEMIG grants APQ-02030-10 and CEX-PPM-00235-12.
This research was performed using the facilities of the Laborat\'orio de Astrof\'isica Computacional da Universidade Federal de Itajub\'a (LAC-UNIFEI).
We employed catalogs from CDS/Simbad (Strasbourg)
and Digitized Sky Survey images from the Space Telescope Science
Institute (US Government grant NAG W-2166) This work has made use of data from the European Space Agency (ESA) mission Gaia (http://www.cosmos.esa.int/gaia), processed
by the Gaia Data Processing and Analysis Consortium (DPAC,
http://www.cosmos.esa.int/web/gaia/dpac/consortium).

%%%%%%%%%%%%%%%%%%%%%%%%%%%%%%%%%%%%%%%%%%%%%%%%%%



\bibliographystyle{mnras}
\bibliography{refs} % if your bibtex file is called example.bib

%\begin{thebibliography}{99}
%\end{thebibliography}

%%%%%%%%%%%%%%%%%%%%%%%%%%%%%%%%%%%%%%%%%%%%%%%%%%

%%%%%%%%%%%%%%%%% APPENDICES %%%%%%%%%%%%%%%%%%%%%

%\appendix

% \section{Some extra material}

% If you want to present additional material which would interrupt the flow of the main paper,
% it can be placed in an Appendix which appears after the list of references.

%%%%%%%%%%%%%%%%%%%%%%%%%%%%%%%%%%%%%%%%%%%%%%%%%%

% Don't change these lines
\bsp	% typesetting comment
\label{lastpage}
\end{document}